\documentclass[journal,twoside,web]{IEEEtran}
\usepackage{tmi}
\usepackage{cite}
\usepackage{amsmath,amssymb,amsfonts}
\usepackage{graphicx}
\usepackage{textcomp}

\usepackage{hyperref}
\usepackage[binary-units=true]{siunitx}
\usepackage{mathtools}
\usepackage{commath}     % ode
\usepackage[T1]{fontenc}
\usepackage{algorithm}
\usepackage{algpseudocode}
\usepackage{stfloats}
\usepackage{booktabs}
\usepackage{tabularx}

\usepackage{mydefs}

\begin{document}
\bstctlcite{IEEEexample:BSTcontrol}

\title{Joint Design of RF and gradient waveforms via~auto-differentiation\\
for 3D tailored excitation in MRI}

\author{Tianrui Luo,
Douglas C. Noll,
\IEEEmembership{Senior Member, IEEE},
Jeffrey A. Fessler,
\IEEEmembership{Fellow, IEEE},
Jon-Fredrik Nielsen
  \thanks{This work was supported by the National Institutes of Health under
  Grants R01EB023618, R21AG061839.}
  \thanks{T. Luo, D. Noll, and J. -F. Nielsen are with the Department of
  Biomedical Engineering,
  University of Michigan, Ann Arbor, MI 48109 USA (e-mail:
  tianrluo@umich.edu, dnoll@umich.edu, jfnielse@umich.edu).}
  \thanks{J. Fessler is with the Department of Electrical Engineering and
  Computer Science, University of Michigan, Ann Arbor, MI 48109 USA (email:
  fessler@umich.edu).}
  }

\maketitle

% Just a quick link to the Author's guideline:
% \url{https://www.embs.org/tmi/authors-instructions/}

\begin{abstract} % 250 words limit
  This paper proposes a new method for joint design of radiofrequency (RF) and
  gradient waveforms in Magnetic Resonance Imaging (MRI), and applies it to the
  design of 3D spatially tailored saturation and inversion pulses.
  The joint design of both waveforms is characterized by the ODE Bloch
  equations, to which there is no known direct solution.
  Existing approaches therefore typically rely on simplified problem
  formulations based on, e.g., the small-tip approximation or constraining the
  gradient waveforms to particular shapes, and often apply only to specific
  objective functions for a narrow set of design goals (e.g., ignoring hardware
  constraints).
  This paper develops and exploits an auto-differentiable Bloch simulator
  to directly compute Jacobians of the (Bloch-simulated) excitation pattern with
  respect to RF and gradient waveforms.
  This approach is compatible with \emph{arbitrary} sub-differentiable loss
  functions, and optimizes the RF and gradients directly without restricting the
  waveform shapes.
  For computational efficiency, we derive and implement explicit Bloch simulator
  Jacobians (approximately halving computation time and memory usage).
  To enforce hardware limits (peak RF, gradient, and slew rate), we use a change
  of variables that makes the 3D pulse design problem effectively unconstrained;
  we then optimize the resulting problem directly using the proposed
  auto-differentiation framework.
  We demonstrate our approach with two kinds of 3D excitation pulses that cannot
  be easily designed with conventional approaches:
  Outer-volume saturation (\ang{90} flip angle),
  and inner-volume inversion.
  % MRI experiments with these pulses produced excitation patterns having NRMSE
  % of 4.3\% and 6.6\% for a simple cuboid IV shape, for OV saturation and IV
  % inversion, respectively, and NRMSE of 20\% for an inversion pulse designed
  % for a more complex 3D IV shape.
  % Compared to a non-joint design that only optimizes the RF pulse, our joint
  % approach improves excitation performance significantly (e.g., 37\% reduction
  % in NRMSE in one simulated excitation experiment).
\end{abstract}

\begin{IEEEkeywords}
  Auto-differentiable Bloch simulator, Constrained joint pulse design,
  Inner-volume inversion, Large flip-angle pulse, Outer-volume saturation,
  Tailored RF pulse design.
\end{IEEEkeywords}

\section{Introduction}
\IEEEPARstart{I}{n} a magnetic resonance imaging (MRI) experiment, the dynamic
system relationship between the applied radiofrequency (RF) and gradient
magnetic fields, and the instantaneous spin magnetization change they induce, is
concisely described by the Bloch equation.
While it is straightforward to calculate the magnetization pattern resulting
from a given set of RF and gradient waveforms and tissue parameters,
\emph{inverting} the Bloch equation to obtain the waveforms that produce a given
desired excitation pattern can be challenging.

This Bloch inversion task is conventionally called an ``RF pulse design''
problem, reflecting the fact that the most common way to design excitation
pulses in MRI is to pre-define the gradients in some way, and then optimize only
the (complex) RF waveform.
Even with that simplification, the design problem remains non-linear and
non-convex.
Another common simplification is to apply the small-tip approximation
\cite{Pauly1989kSpace} that can give reasonable excitation accuracy even for
flip angles as high as \ang{90}, at least for conventional 1D (slice-selective)
excitations where the instantaneous flip angle during RF excitation remains
relatively low.
The small-tip approximation leads to a linear (Fourier) relationship between
applied fields and the resulting magnetization pattern, and provides intuition
about the excitation process by defining an ``excitation k-space'' trajectory
and viewing RF transmission as depositing energy along that trajectory.

The more difficult problem of \emph{jointly} optimizing both RF and gradient
waveforms has been approached in various ways.
Several methods are based on the small-tip approximation, and on optimizing the
gradients over a restricted set of waveform shapes, such as ``spoke'' or
``kt-point'' locations in excitation k-space \cite{Ma2011Joint,
Zelinski2008Sparsity, Cloos2012kTpoints, Yoon2012Fast, Grissom2012Small}
or parameterized echo-planar or non-Cartesian trajectories \cite{Yip2007Joint,
Hardy1988Optimization, Davids2016Fast, Shao2012Advanced, Malik2012SPINS}.
A more general small-tip design approach for 3D tailored excitation used a
B-spline parametrization of the gradient trajectory that is not restricted to
particular fixed waveform shapes \cite{Sun2016Joint}.
These approaches work well for small-tip excitations, but not for applications
such as tailored saturation or inversion.
In addition, even when the final desired flip angle is small, the
\emph{instantaneous} flip angle during RF excitation can be large enough to
violate the small-tip assumption \cite{Sun2015Thesis}.
This model mismatch can cause noticeable differences between the Bloch-simulated
excitation pattern and that predicted by the small-tip model used in the design.

Another limitation of previous approaches is that the design loss functions are
typically limited to certain forms such as least squares (LS) based on the
complex transverse excited magnetization, although adaptations to magnitude
least squares (MLS) costs have been proposed \cite{Setsompop2008Magnitude}.
Adding hardware constraints to the design formulation adds an additional layer
of complexity that is often either ignored during pulse design, or controlled
indirectly via, e.g., Tikhonov regularization of the RF waveform
\cite{Grissom2012Small}.

This work\footnote{Open sourced,
\href{http://github.com/tianrluo/AutoDiffPulses}
{github.com/tianrluo/AutoDiffPulses}
}
approaches the Bloch inversion task in a more direct and general way that is
applicable to the joint design of RF and gradient waveforms for tailored
multi-dimensional excitation in MRI.
We temporally discretize the pulse, assuming piecewise constant gradient and RF
within every time segment.
Our method does not rely on the small-tip approximation, works for arbitrary
sub-differentiable loss functions, and incorporates hardware constraints.
Our approach contains three key elements:
First, we derive analytic expressions for the Jacobian operations needed for the
Bloch inversion for a unit (discrete) time step.
Second, we incorporate these discrete-time Jacobian operations into an automatic
differentiation framework \cite{Paszke2017Automatic}, to obtain the Jacobian
that relates the final magnetization pattern (at the end of the pulse) to the RF
and gradient waveforms.
Third, we enforce hardware limits by a change of variables that makes the
optimization problem effectively unconstrained.

The paper is organized as follows.
Section \ref{sec:thry} gives a general form of the joint design problem, and
derives the explicit Jacobians useful for accelerating the proposed
auto-differentiation pulse design tools.
Sections \ref{sec:meth} and \ref{sec:res} apply our pulse design tool to two
large-tip excitation problems, and validate the results experimentally on a 3T
MRI scanner.
Sections \ref{sec:disc} and \ref{sec:conc} discuss and conclude this work.

\section{Theory}\label{sec:thry}

\subsection{Problem Formation}
We discretize 3D space on a regular grid with a total $n_M$ voxels ("spins").
These spins can have different parameters, e.g. T1, T2, and off-resonance.
Let $n_T$ denote the length (number of time points) of the pulse to be designed.
For (single coil) joint design of complex RF waveform $b\in\mathbb{C}^{n_T}$ and
gradient $g\in\mathbb{R}^{n_T \times 3}$ we are interested in tackling the
following general problem:
\begin{equation}\label{eq:loss}
  \begin{array}{cl}
    \underset{g\in\RR^{n_T\times3},\; b\in\CC^{n_T}}{\argmin}
      & \loss \is f(M_T(g, b), M_D) + \lambda\mathcal{R}       \\
    \text{s.~t.}
      & \|b\|_\infty \le b_{\mathrm{max}}                      \\
      & \|g\|_{\infty, \infty} \le g_{\mathrm{max}}            \\
      & \|\mathsf{D} g\|_{\infty, \infty} \le s_{\mathrm{max}}, \\
  \end{array}
\end{equation}
where $\loss$ is the loss function;
$M_D\in \RR^{n_M\times 3}$ is the target (\emph{D}esired) magnetization
pattern (a 3-dimensional magnetization vector at each spatial location);
$M_T\in \RR^{n_M\times 3}$ is the magnetization at the end of the pulse (time
$T$) obtained by integrating the Bloch equation;
$f$ is the excitation error metric (e.g., a common choice is least-square error
of transverse magnetization, i.e.,
$\|M_T[:,1{:}2] - M_D[:,1{:}2]\|_F^2$);
and $\mathcal{R}$ is an optional regularizer with weight $\lambda$ (a common
choice is $\mathcal{R} = \|b\|_2^2$ to control peak RF amplitudes and SAR
indirectly).
For the constraints, we have
$b_\maxrm$, $g_\maxrm$, and $s_\maxrm$ for peak RF, gradient, and slew rate,
respectively;
$\mathsf{D}\in\mathbb{R}^{n_T\times n_T}$ is the temporal difference matrix
divided by $\delta_t$, i.e., $\mathsf{D}g$ takes the 1st order temporal
derivative of $g$ and yields the slew rate;
and $\|\cdot\|_\infty$, and $\|\cdot\|_{\infty,\infty}$ are entry-wise norm
returning the largest absolute value of the operand elements.

Problem~\eqref{eq:loss} is challenging for two main reasons:
First, the objective is non-convex with respect to its arguments, and is
constrained.
Second, neither $M_T(g,b)$, nor its Jacobians $\partial M_T/\partial g$ and
$\partial M_T/\partial b$ that would be needed to directly minimize~
\eqref{eq:loss}, have an explicit expression in $g$ and $b$.
To the best of our knowledge, existing methods all deal with simplifications of
problem~\eqref{eq:loss} based on, e.g., the small-tip, or spin domain models.
In this work, we minimize~\eqref{eq:loss} directly, and assume only that the
temporal integration of the Bloch equation is well-approximated by a
discrete-time Bloch simulator.

\subsection{Auto-Differentiation}

We propose to compute the necessary derivatives\footnote{Or Clarke generalized
subdifferentials for non-smooth objectives \cite{Clarke1983Nonsmooth}.}
using auto-differentiation \cite{Griewank2008Evaluating}, such that
problem~\eqref{eq:loss} can be optimized for \emph{arbitrary} error metric $f$
and regularization $\mathcal{R}$.
Auto-differentiation tools, e.g., PyTorch \cite{Paszke2017Automatic}, decouples
computations into stages, and constructs the Jacobian operations at each stage.
These single-stage Jacobians are eventually combined using the chain rule.
For instance, with a PyTorch based Bloch simulator that computes $M_T(g, b)$,
one implicitly obtains $\partial M_T/\partial g$ and $\partial M_T/\partial b$.
The loss derivatives with respect to the variables we wish to optimize, i.e.,
$\partial \loss/\partial g$, and  $\partial \loss/\partial b$,
can then be obtained by combining these expressions with
$\partial \loss/\partial M_T$.
This approach allows us to directly optimize $g$ and $b$
with respect to arbitrary losses.

\subsection{Explicit Jacobian Operations}
Auto-differentiation tools provide implicit Jacobian operations (also known as
the default backward operations in auto-differentiation context) formed from
tracking all elementary computations (e.g., addition, multiplication, etc).
Such tools also allow users to substitute default Jacobian operations with their
own implementations.
In practice, such \emph{explicitly} implemented Jacobian operations can be more
efficient both computationally and memory-wise.
Bloch simulation is typically the most computationally expensive stage in
relating pulse waveforms to objective costs.
Having explicit Jacobians of the Bloch simulator can therefore accelerate the
computation.

To derive discrete time ($\delta_t$) explicit Jacobians in the rotating frame,
for all magnetic spins, we assume equilibrium spin magnitudes of $1$, relaxation
constants $e_1\is\exp(-\delta_t/T_1)$, $e_2\is\exp(-\delta_t/T_2)$, and
gyromagnetic ratio $\gamma$.
% note: according to wiki, the SI unit for gyromagnetic ratio is Rad⋅s⁻¹⋅T⁻¹,
% hence no need to introduce \bar{γ} := 2π⋅γ
At time $t$, the rotating frame effective magnetic field (B-effective),
$B_t\in\mathbb{R}^3$, causes the magnetic spin state $m_t\in\mathbb{R}^3$, to
precess (rotate) about an axis $u_t\is B_t/\|B_t\|_2$ by angle
$\phi_t\is-\gamma\delta_t\|B_t\|_2$.
One iteration of discrete time Bloch simulation can be expressed as:
\begin{equation}\label{eq:iterforw}
  m_{t+1} = ER_{t}m_{t} + e,
\end{equation}
where $E\is\diag([e_2,e_2,e_1])$, $e\is\trans{[0,0,1-e_1]}$ model the
relaxations;
$R_t=\cos(\phi_t)I+(1-\cos(\phi_t))u_t\trans{u_t}+\sin(\phi_t)[u_t]_\times$
models the rotation;
$I$ is the 3D identity matrix, $\diag([1, 1, 1])$;
and $[u_t]_\times$ denotes the cross product matrix of $u_t$, i.e.,
$[u_t]_\times m_t = u_t \times m_t$.
The rotation matrix $R_t$ is spatially dependent, as it accounts for B-effective
which incorporates applied gradients, off-resonance, etc.
Relaxation terms, as they depend on the underlying tissue property, are
generally also spatially dependent.
To avoid notation clutter, we have not indicated those spatial dependencies in
Eq. \eqref{eq:iterforw};
rather, Eq. \eqref{eq:iterforw} can be considered to hold for a single spin
isochromat, with the appropriate $R_t$ and relaxation terms for that
isochromat.

One can verify the following recursive expressions for partial derivatives of
the loss with respect to $m_t$ and $B_t$:
\begin{equation}\label{eq:iterback}
  \begin{aligned}
    \pd{\loss}{m_t} & = \trans{R_{t}}E\pd{\loss}{m_{t+1}}\eqqcolon h_{t},    \\
    \pd{\loss}{B_t} & = \gamma\delta_t/\phi_t(u_t\trans{u_t}-I)\pd{\loss}{u_t}
                        -\gamma\delta_t\pd{\loss}{\phi_t}u_t,                \\
    \pd{\loss}{u_t} & = \phi_t\left(\mathfrak{c}_t(m_t\trans{u_t}
                                                   +\trans{m_t}u_t I)
                                    +\mathfrak{s}_t[m_{t}]_\times
                              \right) Eh_{t+1},                               \\
    \pd{\loss}{\phi_t} & = \trans{([u_t]_\times R_{t}m_{t})} Eh_{t+1},
  \end{aligned}
\end{equation}
where $\mathfrak{c}_t\is\left(1-\cos(\phi_t)\right)\!/\phi_t$,
$\mathfrak{s}_t\is\sin(\phi_t)/\phi_t$.
Given $\partial B_t/\partial g_t$ and $\partial B_t/\partial b_t$
(which are easy to compute),
we obtain the necessary derivatives for the
joint optimization by the chain rule:
\begin{equation*}
  \pd{\loss}{g_t}=\pd{\loss}{B_t}\cdot\pd{B_t}{g_t}, \quad
  \pd{\loss}{b_t}=\pd{\loss}{B_t}\cdot\pd{B_t}{b_t}.
\end{equation*}

Using the explicit Jacobians in \eqref{eq:iterback} for the Bloch simulator
operations halved both the computation time and memory use compared to the
default implicit Jacobian operations provided by PyTorch (v1.3).

The remaining Jacobians, such as $\partial \loss/\partial M_T$,
$\partial B_t/\partial b_t$, and $\partial B_t/\partial g_t$, typically do
not involve complicated computations.
Also, they can vary with different objectives, e.g., switching from LS to MLS;
or with different excitation settings, e.g., uniform vs non-uniform transmit
sensitivities.
For program generality, we left these remaining Jacobians to be obtained
implicitly by the auto-differentiation framework.

\begin{figure}
  \centering
  \includegraphics[width=0.6\columnwidth]{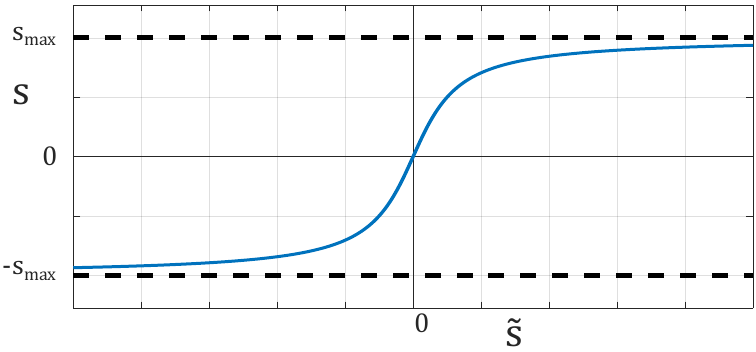}
  \vspace{-0.5\baselineskip}
  \caption{Turning constrained slew rate $s$ into unconstrained $\tilde{s}$, by
    change of variable tan$^{-1}$.}
  \vspace{-\baselineskip}
  \label{fig:arctan}
\end{figure}

\subsection{Constraints}\label{sec:constraints}
Constrained optimization often requires extra effort to ensure solution
feasibility,
such as feasible set projection and constraint substitution with penalizations.
This would involve crafting projection algorithms, and tuning penalty
parameters.
For problem~\eqref{eq:loss}, in the absence of convexity,
we use a change of variables \cite{Sisser1981Elimination,
Boyd2004ConvexOptimization} that converts the problem into an effectively
unconstrained one and avoids such extra effort during optimization.

Let $s\in\RR^{n_T \times 3}$ denote the slew rate, i.e., $s=\mathsf{D}g$.
Define $\tilde{s}=\tan(\pi/2\cdot s/s_\maxrm)$;
$\tilde{\rho}\is\tan(\pi/2\cdot|b|/b_\maxrm)$, $\theta\is\angle b$.
We automatically have $\|b\|_\infty \le b_\maxrm$ and
$\|s\|_{\infty,\infty} \le s_\maxrm$ always satisfied (Fig. \ref{fig:arctan}).
Thus, we reformulate problem~\eqref{eq:loss} as:
\begin{equation}\label{eq:lossalter}
  \begin{array}{cl}
    \underset{\tilde{s}\in\RR^{n_T\times3};\;\tilde{\rho},\;\theta\in\RR^{n_T}}
      {\argmin}
       & \mathcal{L} \coloneqq f(M_T(g, b), M_D) + \lambda\mathcal{R}   \\
    \textrm{s. t.}
       & \|g\|_{\infty, \infty} \le g_{\mathrm{max}}                    \\
       & \mathsf{D}g=2s_\mathrm{max}/\pi \cdot \tan^{-1}(\tilde{s})     \\
       & b=2b_\maxrm/\pi \cdot \exp(\iota\theta)\tan^{-1}(\tilde{\rho}). \\
  \end{array}
\end{equation}

In practice, for change of variable, $\tan^{-1}$ can be replaced with any other
\emph{strictly monotone} function, e.g., sigmoid, that maps an unconstrained
domain to an interval.
% Such functions have non-zero derivatives everywhere
% and do not introduce new local minima
% to the already non-convex problem \eqref{eq:loss}.

Empirically, for 3D tailored pulse design, we observe that, with extended
kt-points initializations \cite{Sun2016Joint},
gradient amplitudes are well below typical max gradient constraints
(\SI{5}{\gauss\per\centi\meter})
prior to and throughout the optimization procedure.
Hence, while problem~\eqref{eq:lossalter} is still constrained formally, its
max gradient is practically inactive.
We thus treated it as an unconstrained problem for the results shown in this
paper.

\subsection{Optimization Algorithm}
We select initial waveforms $g$ and $b$ that satisfy the constraints.
To minimize \eqref{eq:lossalter},
we alternatingly update $\tilde{\rho}$, $\theta$, and $\tilde{s}$, as shown in
Algorithm~\ref{alg:AlterMin}.
This alternating strategy is commonly used in existing joint design approaches
\cite{Sun2016Joint, Yip2007Joint}, and helps reduce the problem size for the
L-BFGS algorithm used in updating the pulse.
With auto-differentiation, the optimization algorithm can be formulated without
reference to the specific loss function, as demonstrated with very different
design problems in section~\ref{sec:meth}.
We use the L-BFGS optimizer provided by PyTorch for updating the variables
within an iteration.
The number of iterations may depend on pulse initializations.
We empirically choose $N=10$ for experiments in this work.
This choice can vary with applications.

\begin{algorithm}[H]
  \caption{Alternating Minimization} \label{alg:AlterMin}
  \begin{algorithmic}[1]
    \INPUT Variables: $g$, $b$; Number of iterations: $N$
    \State Compute $\tilde{\rho}$, $\theta$ and $\tilde{s}$
      from $g$, $b$
    \For{$n$ = 1 to $N$}
      \State Fix $\tilde{s}$; Optimize $\tilde{\rho}$, $\theta$, using L-BFGS
      \State Fix $\tilde{\rho}$, $\theta$; Optimize $\tilde{s}$, using L-BFGS
    \EndFor
    \State Compute $g$ and $b$, from $\tilde{\rho}$, $\theta$ and $\tilde{s}$
    \State \Return $g$, $b$
  \end{algorithmic}
\end{algorithm}
\vspace{-1.5\baselineskip}

\section{Methods}\label{sec:meth}
To demonstrate the utility and generality of our approach, we designed two
different kinds of 3D tailored pulses: outer-volume (OV) saturation, and
inner-volume (IV) inversion.

\subsection{3D Outer-Volume Saturation Pulse Design}
Outer-volume saturation pulses can be used to limit the imaging field of view
(FOV), and hence has the potential to reduce both the time needed for data
acquisition as well as motion artifacts from, e.g., the chest wall or abdomen in
body imaging applications \cite{Luo20193D, Mitsouras2005Strategies,
Wilm2007Reduced}.
OV saturation pulses should ideally have a high flip angle in the OV region
(e.g., 90 degrees), while leaving the IV unperturbed.
These pulses are typically followed immediately by a gradient crusher.
Since the phase of OV magnetization prior to the crusher is unimportant, we use
MLS loss in design, and include a regularization term on RF power to indirectly
control SAR as well as to demonstrate the generality of our approach:
\begin{equation}\label{eq:loss_OV90}
  \loss_{90} = \||M_T[:,1{:}2]|-|M_D[:,1{:}2]|\|_2^2 + \lambda\|b\|_2^2,
\end{equation}
where, $|M[:,1{:}2]|\is\mathrm{abs}(M[:,1]+\iota M[:,2])$, is a vector function
computing magnitudes of spin transverse magnetizations.
For the target excitation profile, we set rows in $M_D$ to $[1,0,0]$ for OV
spins, and $[0,0,1]$ for IV spins.
We implemented this loss in PyTorch to obtain the Jacobian
$\partial \loss_{90}/\partial M_T$ as described in the Theory section.

In principle, small-tip based 3D tailored design approaches can also be applied
to this loss by scaling the designed RF pulse to attain the desired \ang{90}
flip (although the resulting pulse may exceed peak RF limits).
We therefore compare our approach with the small-tip method in
\cite{Sun2016Joint}, starting with the same initial $b$ and $g$ waveforms in
both cases (initialized as described in \ref{sec:meth_init}).

\subsection{Inner-Volume Inversion Pulse Design}

Next we designed another type of excitation pulse that is difficult to design
using conventional approaches: an IV \emph{inversion} pulse.
Such a pulse may be useful for, e.g., selective inversion of arterial blood for
flow territory mapping in perfusion imaging.
For this pulse we propose a very different excitation loss based on the
\emph{longitudinal} magnetization:
\begin{equation}\label{eq:loss_IV180}
  \loss_{180} = \|M_T[:,3]-M_D[:,3]\|_2^2 + \lambda\|b\|_2^2.
\end{equation}
We set rows in $M_D$ to $[0,0,1]$ for OV spins, and $[0,0,-1]$ for IV spins.
We also implement this loss in PyTorch.

\begin{figure*}[!b]
  \centering
  \includegraphics[width=0.7\textwidth]{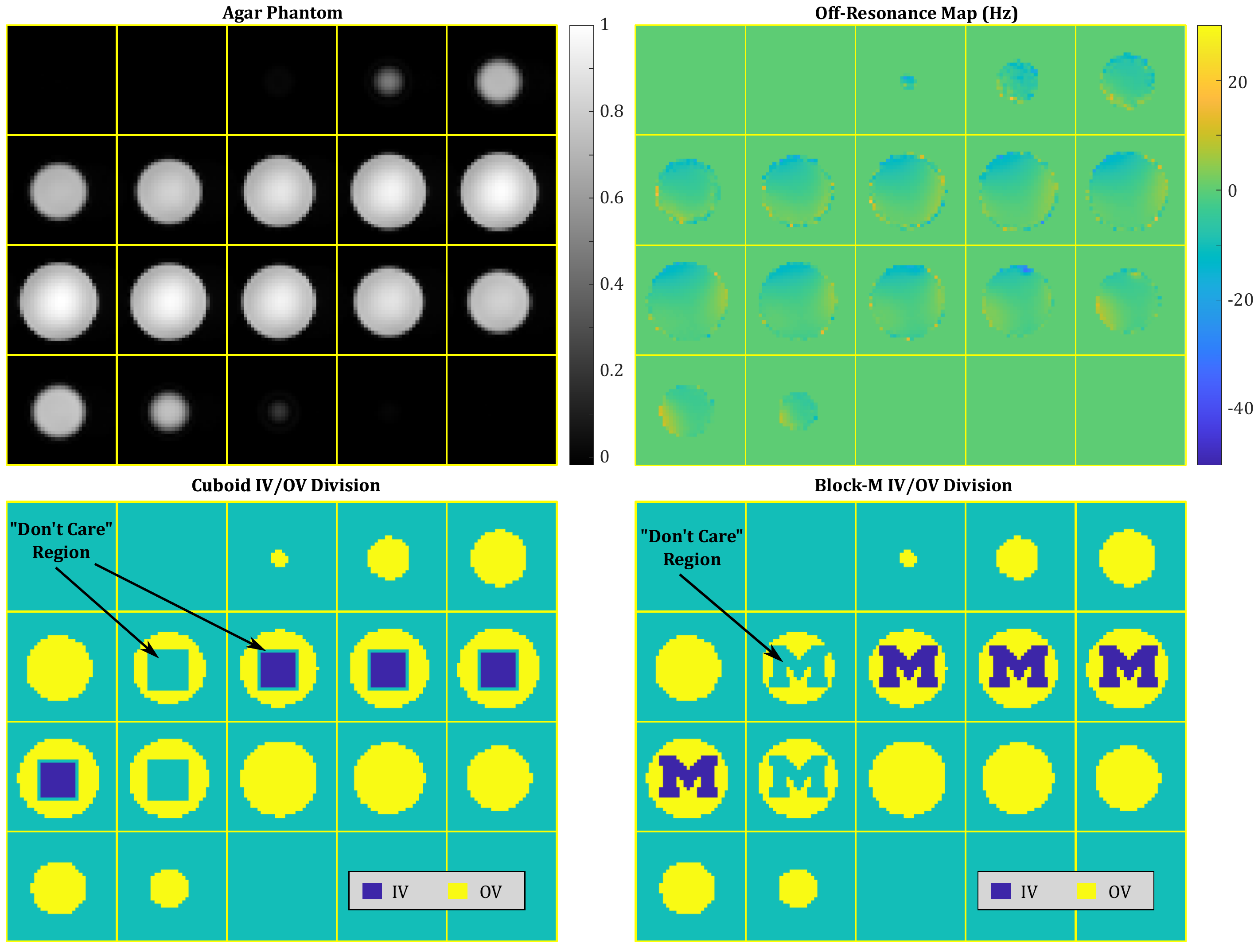}
  \vspace{-0.5\baselineskip}
  \caption{Experimental phantom, and the two target patterns (IV/OV divisions)
    used in our experiments.
    Top left: Magnitude image of the uniform Agar phantom.
    Top right: Observed field map, used in the pulse design to account for B0
    inhomogeneity.
    A conservative mask that is 1-voxel-wide larger than the phantom support was
    used to ensure that the phantom boundary was included in the design.
    This expanded mask is the likely cause for the relatively large B0 values in
    some pixels at the edge of the mask (that are likely just outside the
    phantom).
    Bottom: Cuboid (left) and ``block-M'' (right) target patterns.
    We prescribed a ``don't care'' (region with arrows) at the boundary between
    the IV and OV regions that is excluded when calculating the design loss.
    For the cuboid pattern, the don't care region included the entire 3D IV/OV
    boundary, whereas for the block-M pattern, only the top and bottom slices
    (slices 6 and 11;  slice numbers increase left-to-right and top-to-bottom)
    were included due to the low in-plane spatial resolution of the design grid.
  }
  \vspace{-\baselineskip}
  \label{fig:setup}
\end{figure*}

\subsection{Pulse Initializations}\label{sec:meth_init}

The loss in problem \eqref{eq:lossalter} is non-convex, and the choice of
initial $g$ and $b$ waveforms influences the final excitation result.
How best to initialize these waveforms is an open problem.
In \cite{Sun2016Joint}, the initialization problem (in the context of small-tip
3D tailored excitation) was addressed by evaluating two popular choices for the
excitation k-space trajectory, stack-of-spirals and SPINS \cite{Malik2012SPINS},
along with a novel alternative approach, ``extended kt-points'', that chooses
gradients based on the desired (target) excitation pattern.
Sun \emph{et al}. showed that the extended kt-points approach produces
comparable or better excitation accuracy than stack-of-spirals or SPINS
\cite{Sun2016Joint}, so we chose it for the experiments in this work.

Once the gradients were initialized in this way, we initialized $b$ using the
approach in \cite{Yip2005Iterative}.
These initial RF waveforms were scaled down when necessary to satisfy the
$b_\maxrm$ constraint.

%% skipping description about \lambda, as it was not discussed in Sun2016Joint
% \subsection{Regularization Weighting}
% Conventionally, the weighting coefficient $\lambda$ for regularizing RF power,
% $\|b\|_2^2$, has been used to indirectly control RF peak amplitude, etc.
% In this work, peak RF constraint is directly enforced by change of variables.
% We keep this regularization term with $\lambda=4$ in all experiments only for
% demonstration purposes, and do not claim optimality.

\subsection{B-effective Computation}
The particular form of B-effective depends on the excitation objective and other
application-specific components.
Besides the RF and gradient waveforms, it often also contains an off-resonance
map (that may vary with time) and transmit sensitivity maps.
Other factors such as gradient non-linearity can also be included in
B-effective.
For the single transmit coil phantom studies in this work, B-effective accounts
for RF, gradient, and a static off-resonance map.
Specifically, at time $t$, let $b_t \in \mathbb{C} $ and $g_t \in \mathbb{R}^3$
denote the instantaneous RF and gradients, respectively.
For position $r$ relative to the scanner iso-center, with off-resonance
$\omega(r)$, the instantaneous B-effective is:
\begin{equation}\label{eq:Beff}
  B_t = [\mathfrak{R}(b_t), \mathfrak{I}(b_t), \inp{g_t}{r} + \omega(r)/\gamma],
\end{equation}
where $\mathfrak{R}$ and $\mathfrak{I}$ extract the real and imaginary component,
respectively.

\begin{table}[t]
  %\vspace{-\baselineskip}
  \caption{Pulse Duration and TR/TE}
  \label{tbl:timing}
  \centering
  \renewcommand{\arraystretch}{1}
  \begin{tabularx}{\columnwidth}{CCCC}
    \toprule
    Parameters      & OV90                      & IV180
      & IV180M        \\
    \midrule
    Pulse Duration  & \SI{6.5}{\ms}             & \SI{4.8}{\ms}
      & \SI{4.5}{\ms} \\
    TR / TE         & \SI{2}{\s} / \SI{15}{\ms} & \multicolumn{2}{c}{\SI{3}{\s}
      / minimum TE}   \\
    \bottomrule
  \end{tabularx}
  \vspace{-1.5\baselineskip}
\end{table}

\subsection{Phantom Experiments}

We performed validation experiments in an Agar phantom on a GE MR750 3T scanner.
Fig. \ref{fig:setup} illustrates the experimental setup, including the
prescribed IV and OV regions.
All experiments used the same observed off-resonance map in the pulse design
(Fig. \ref{fig:setup}).
We used $T_1=\SI{1.47}{\s}$, $T_2=\SI{70}{\ms}$ in the Bloch simulation during
designs.
For all studies, we conducted the 3D design on a \num{32 x 32 x 20} voxel grid
of FOV \SI[product-units=power]{24 x 24 x 24}{\cm};
with RF power weighting coefficient $\lambda=4$, and constraints:
$b_\mathrm{max}=\SI{0.25}{\gauss}$,
$g_\mathrm{max}=\SI{5}{\gauss\per\cm}$,
$s_\mathrm{max}=\SI{12}{\gauss\per\cm\per\ms}$.
We quantified excitation performance in simulations with normalized root mean
squared error (NRMSE).
Spins in ``don't care'' regions (Fig. \ref{fig:setup}) were excluded when
calculating the NRMSE.
We ran our design programs on an NVidia 2080 Ti graphics card.
With the settings above, our method uses around \SI{1.1}{\giga\byte} GPU RAM for
all 3 designs.
This includes the intrinsic GPU RAM usage of the PyTorch environment.

We performed three different experiments:
1. OV \ang{90} excitation using the cuboid target pattern shown in
Fig.~\ref{fig:setup} (OV90);
2. IV inversion using that same cuboid target pattern (IV180); and
3. IV inversion with a block-M target pattern (IV180M).
The experiments were implemented using a vendor-agnostic platform for rapid
prototyping of MR pulse sequences \cite{Nielsen2018TOPPE,Layton2017Pulseq}.
IV dimensions were \SI[product-units=power]{9x9x4.8}{\cm} and
\SI[product-units=power]{9x12.8x4.8}{\cm} for the cuboid and block-M target
patterns, respectively.
% We prescribed ``dont't care'' regions (1 voxel wide in the low-resolution
% design grid) along the IV/OV boundary allow the magnetization to change
% continuously across that boundary.
We used a single channel transmit/receive birdcage coil for all experiments and
assumed uniform RF transmit sensitivity during pulse design.
% For IV180M design, due to limited in-plane resolution, we only prescribe
% ``don't care'' regions in z-direction.
To mitigate Gibbs ringing artifacts, we acquired the phantom images at a matrix
size of \num{120x120x48} and then downsized in image space to match the design
grid size \num{32x32x20}.

We used long TR to wait for spin full recovery from saturation and inversion.
For the OV90 experiment, as a substantial volume of the phantom is excited with
large angle, we used TE=\SI{15}{\ms} to intentionally decay signal intensity and
avoid saturating amplifiers in signal receiver during acquisition.
We use minimum TE for the inversion experiments.

\begin{figure}[t]
  \centering
  \includegraphics[width=0.8\columnwidth]{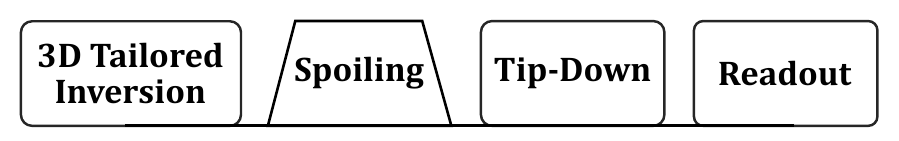}
  \vspace{-0.5\baselineskip}
  \caption{Schematic diagram of the imaging sequence used to characterize
    (validate) the 3D tailored inversion pulses.}
  \vspace{-\baselineskip}
  \label{fig:IV_seq}
\end{figure}

For inversion performance validation, we use the sequence in
Fig.~\ref{fig:IV_seq} to obtain both phase and magnitude phantom images, with
tip-down angle set to \ang{10}.
We expect a $\pi$ phase difference between inverted (IV)  non-inverted (OV)
regions, as the excitation pulse should tip inverted and non-inverted spins in
opposite directions.
In addition to the IV180 and IV180M excitation pulses, we imaged the phantom
using the same sequence settings (TE/TR, flip angle, matrix size) using a
conventional slab-selective Shinnar-Le Roux (SLR) pulse.
We normalized the inversion images using this ``non-inversion'' image to
eliminate receive coil sensitivity weighting (both magnitude and phase) in the
inversion images.
For completeness, the unnormalized images are shown in supplemental materials
(Fig. S5).

\section{Results}\label{sec:res}

\subsection{OV90}

\begin{figure*}[!t]
  \centering
  \includegraphics[width=\linewidth]{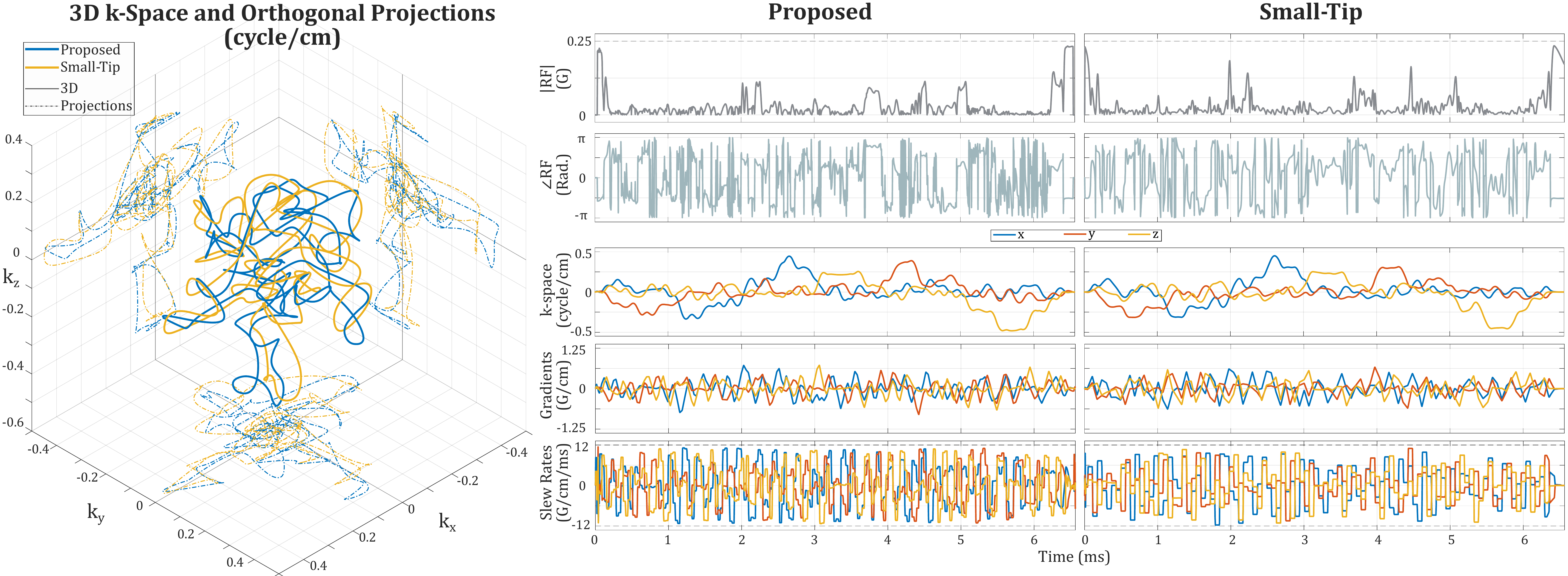}
  \vspace{-0.5\baselineskip}
  \caption{OV saturation pulses for the cuboid IV (Fig.~\ref{fig:setup}),
    designed with our approach (Proposed) and the small-tip method in
    \cite{Sun2016Joint} (experiment OV90).
    The left panel shows the 3D k-space trajectories and their orthogonal
    projections:
    The two trajectories explore largely overlapping regions in excitation
    k-space.
    The right two panels show RF, gradient, and slew rate waveforms.
    Both designs satisfy the constraints, but for the small-tip design it was
    necessary to apply the VERSE \cite{Conolly1988Variable} algorithm near the
    end of the pulse (see Discussion).
    Gradient peak amplitudes remain quite small
    ($\ll\SI{5}{\gauss\per\cm}$), whereas the gradient slew rates are
    frequently near their limit.}
  \vspace{-\baselineskip}
  \label{fig:OV90_pulse}
\end{figure*}

\begin{figure*}[!b]
  \centering
  \includegraphics[width=\linewidth]{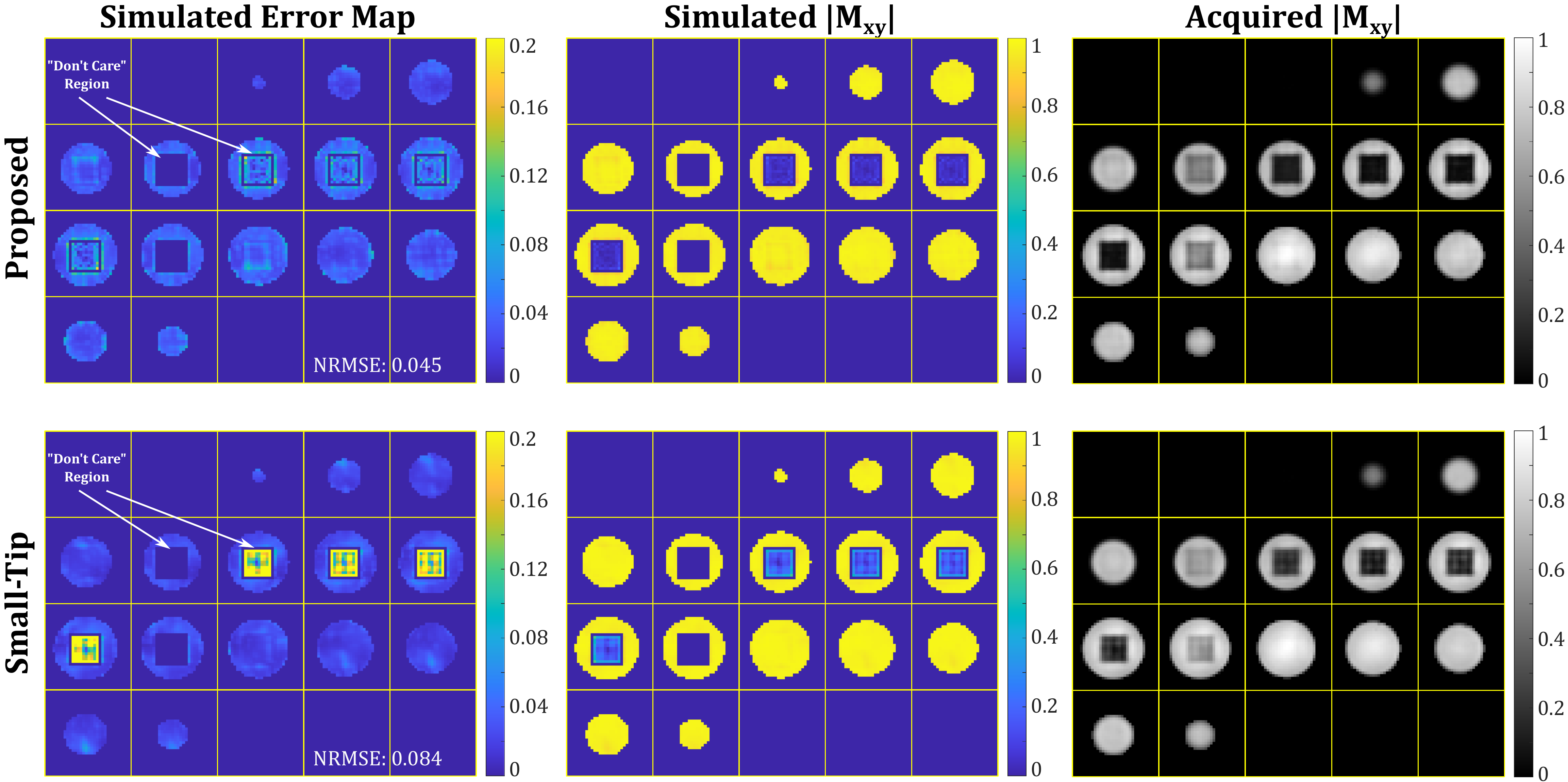}
  \vspace{-0.5\baselineskip}
  \caption{Experimental validation of the pulses shown in
    Fig.~\ref{fig:OV90_pulse}.
    The left panel shows the error map from simulation.
    Our approach has much smaller (-46\%) NRMSE in simulation compared to
    Small-Tip.
    Acquired results (right) agree with the simulations (middle).
    Small-Tip approach has larger error inside the IV:
    This is expected, as the method produces only small-tip pulses, that we then
    scaled to meet the large-tip objective.
    The scaling increases excitation error inside IV while reducing error in the
    OV.
    Our approach directly designs large-tip pulses without this type of
    `scaling' error.
  }
  \vspace{-\baselineskip}
  \label{fig:OV90_result}
\end{figure*}

Figure \ref{fig:OV90_pulse} shows the OV saturation pulses obtained with the
proposed method and the small-tip approach in~\cite{Sun2016Joint},
and Fig.~\ref{fig:OV90_result} shows the corresponding phantom imaging results.
To keep the small-tip RF pulse within peak amplitude limits, we applied VERSE
\cite{Conolly1988Variable} near the end of the pulse.
Our approach required \SI{10}{\min} for design, longer than the small-tip
approach (\SI{2}{\min}).
While the RF waveforms differ markedly, the (excitation) k-space trajectories
are more similar, though differences are clearly observed in the 3D trajectory
plot (Fig.~\ref{fig:OV90_pulse}).

We observe excellent agreement between simulated and acquired excitation
patterns (Fig.~\ref{fig:OV90_result}).
Also, the proposed method produces much lower excitation error than the
small-tip design (46\% lower NRMSE error overall);
this is expected as the small-tip assumption is violated after scaling the RF
to attain the desired \ang{90} flip angle in the OV, which reduces the
error in the OV at the expense of increased error in the IV.

\begin{figure*}[!t]
  \centering
  \includegraphics[width=\linewidth]{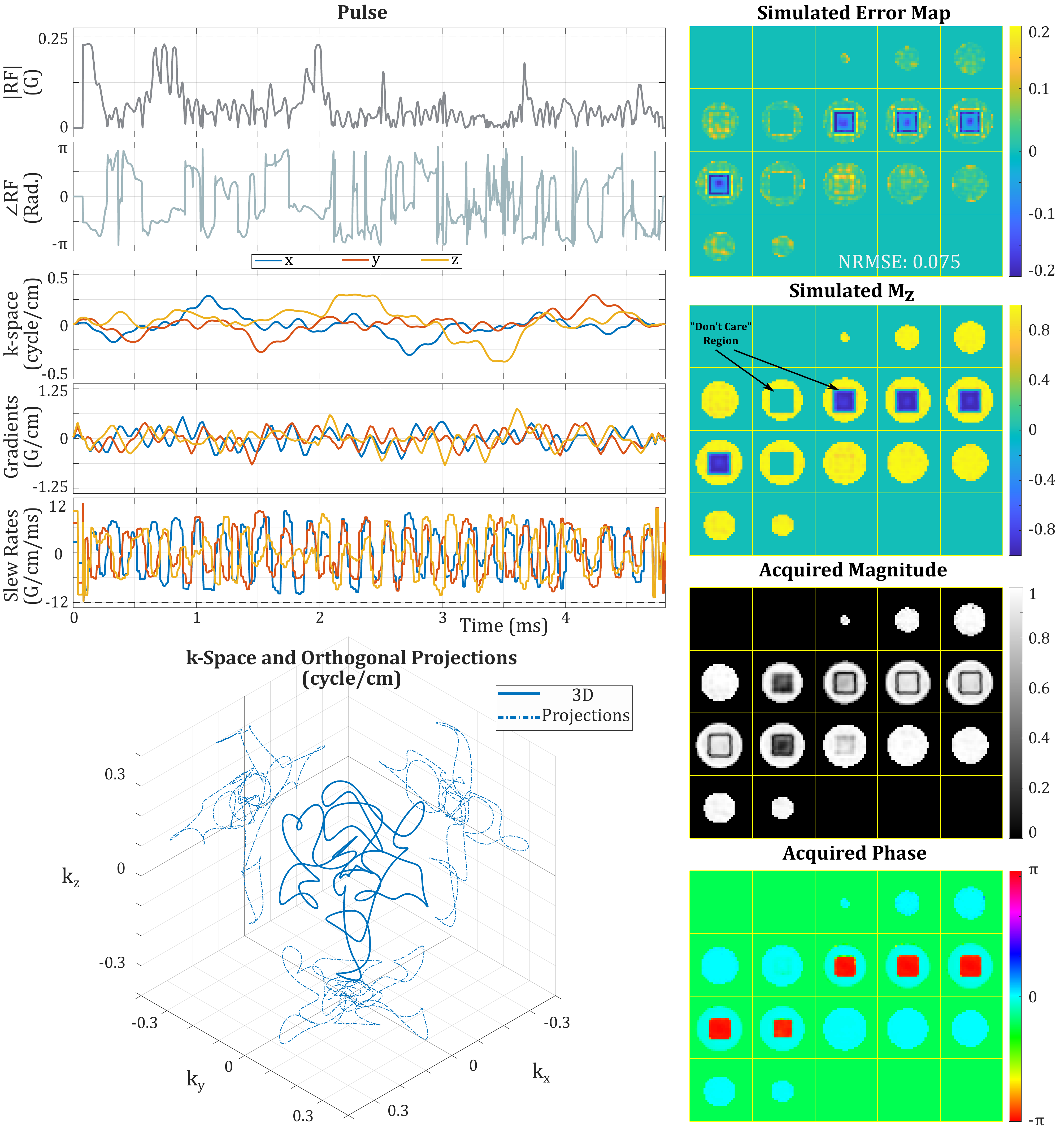}
  \vspace{-\baselineskip}
  \caption{IV inversion results for the cuboid IV pattern (Experiment IV180).
    As desired, the \SI{4.2}{\ms} pulse satisfies all constraints.
    The gradient waveform is again far from its peak constraint of
    \SI{5}{\gauss\per\cm}.
    Compared to the OV90 experiment, the pulse has more extreme slew rate
    waveforms.
    The acquired magnitude and phase (i.e., the ``observed inversion'') were
    obtained with the sequence in Fig.~\ref{fig:IV_seq}.
    We observe good agreement between simulated and acquired inversion patterns.
    The designed pulse successfully inverts the IV, as indicated by similar
    magnitude image intensity in the IV and OV regions (apart from
    transmit/receive coil shading) and a $\pi$ phase shift across the IV/OV
    boundary.
    The dark bands in the acquired images at the IV/OV boundary are due to spin
    saturation from incomplete inversion (and overlap substantially with the
    prescribed ``don't care'' region).
    % The smooth phase variation on the phase image is possibly due to
    % conductivity variations within the phantom.
  }
  \vspace{-\baselineskip}
  \label{fig:IV180_result}
\end{figure*}

\subsection{IV180}\label{txt:IV180_result}

Fig. \ref{fig:IV180_result} shows the results of the cuboid IV inversion
experiment.
Pulse waveforms and images from simulation and phantom experiments are shown.
Pulse design took \SI{6}{\min}.
% The RF amplitudes are higher when k-space waveforms are close to 0, indicating
% a large portion of RF energy is put near the center of k-space.
% Slew rates are more extreme compared to that of the longer OV90 pulse.
% This is possibly partially due to a shorter pulse duration.
Simulations and acquired images are in excellent agreement, and indicate
successful inversion within the IV with errors mainly located at the IV/OV
boundary as expected.
In particular, we observe dark bands along the IV/OV boundary in the magnitude
image.
Spins in this region are not fully inverted, resulting in low signal intensities
in the magnitude image.
The phase image shows an abrupt $\pi$ transition at the IV/OV boundary,
indicating successful IV inversion.

\begin{figure*}[!t]
  \centering
  \includegraphics[width=\linewidth]{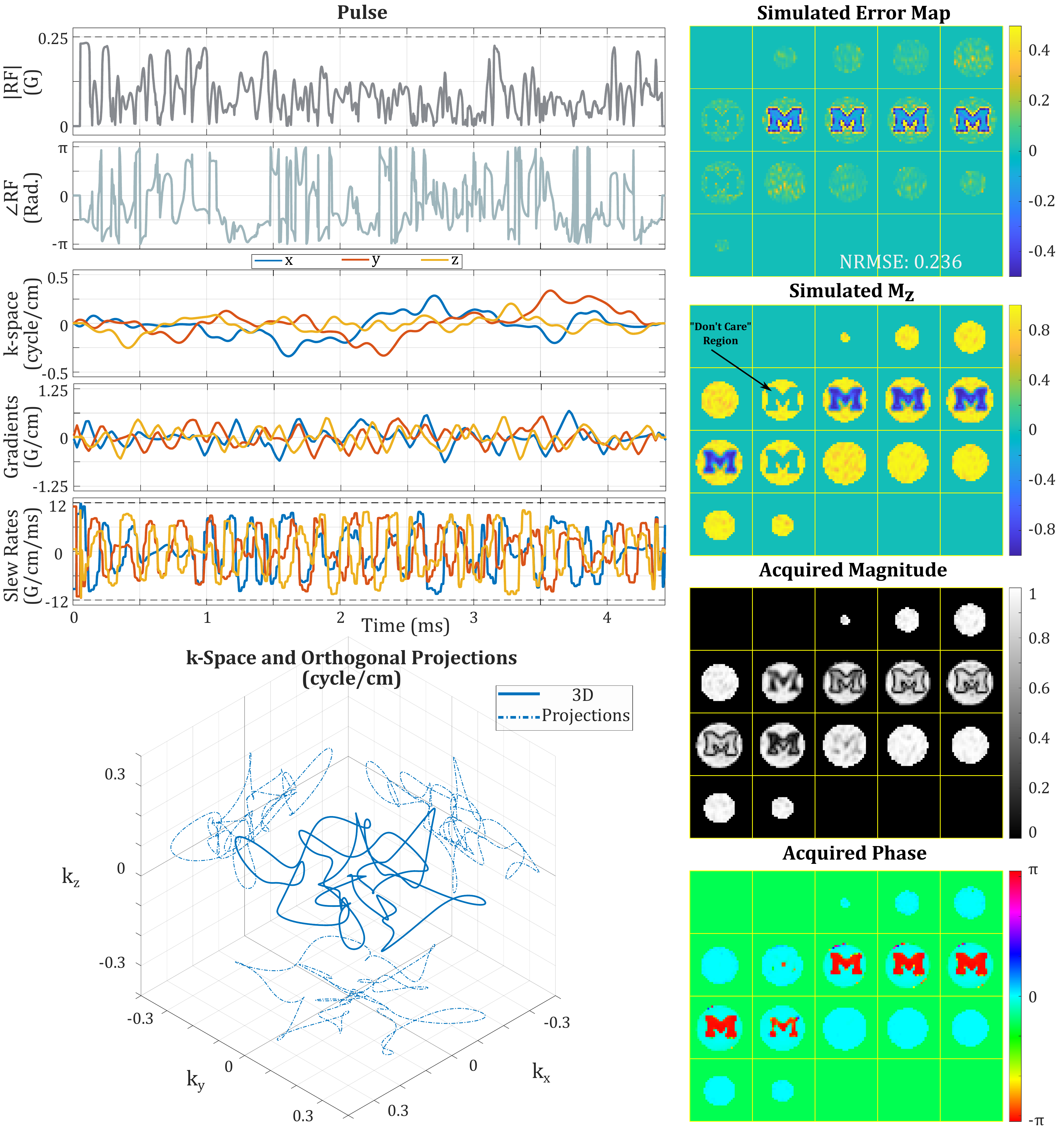}
  \vspace{-\baselineskip}
  \caption{IV inversion results for the block-M target pattern (Experiment
    IV180M).
    The \SI{4.5}{\ms} pulse satisfies all constraints.
    The gradient waveforms are again well below the peak amplitude constraint of
    \SI{5}{\gauss\per\cm}, and slew rates are near the constraint for significant
    portions of the waveform duration.
    As in Fig.~\ref{fig:IV180_result}, the pulse successfully inverts the IV.
    The dark bands in the acquired magnitude image at the IV/OV boundary are due
    to saturation effects arising from the finite resolution (excitation k-space
    extent) of the pulse -- even though only slices 6 and 11 were included in
    the ``don't care'' region in the design due to the low in-plane spatial
    resolution of the design grid (see Fig.~\ref{fig:setup}).
  }
  \vspace{-\baselineskip}
  \label{fig:IV180M_result}
\end{figure*}

\subsection{IV180M}

Fig. \ref{fig:IV180M_result} shows the results of the block-M IV inversion
experiment.
Pulse design took \SI{6}{\min}.
Slew rates are near the limit, similar to the IV180 experiment.
Simulation and acquired images again indicate successful inversion.
The NRMSE is larger than in the IV180 experiment, suggesting a trade-off between
geometry complexity and excitation accuracy.
Excitation error is largest near the in-plane edge of the block-M, where target
$M_z$ changes sharply from \num{1} to \num{-1}.
We again observe dark bands along the IV/OV boundary, and an abrupt phase change
across that boundary, as expected.

\section{Discussion}\label{sec:disc}

We have demonstrated a new approach to joint multi-dimensional excitation pulse
design that directly optimizes both RF and gradient waveforms.
Our approach is not limited to small-tip design problems, and is compatible with
quite general loss/design functions such as those that involve longitudinal
and/or magnitude magnetization.
We validated our approach with 3D tailored large-tip objectives.
For this type of application, the ``extended kt-points'' small-tip
initialization~\cite{Sun2016Joint} led to excellent large-tip results.

We chose to implement our auto-differentiable Bloch simulator with B-effective
as its input for its generality: one can possibly prepend to it arbitrary
functions that compute B-effective from various parameters, such as multi-coil
parallel transmit (pTx) sensitivities, spin movements, and even non-linear
response of gradient amplifiers, gradient delays, etc.
This choice that favors generality may require more memory than software designs
that take RF and gradients as inputs directly, and may require more expensive
hardware with adequate memory for high-dimensional design problems.
In particular, an interface that uses RF, gradient and spin location inputs
requires a memory size proportional to $(N_T+N_T\times 3+N_M\times 3)$, whereas
our interface requires memory proportional to $(N_M \times 3 \times N_T + N_M
\times 3)$.
Our implementation can find use in different scenarios for proof-of-principle
designs that one could then follow by customized simulators that meet specific
computational requirements.

For the Bloch simulator, one may alternatively consider using the hard pulse
approximation, which splits the instantaneous rotation matrix $R_t$ into two
rotations: RF rotation, and transversal rotation due to the applied gradients
and off-resonance.
The hard pulse approximation is the basis for the SLR pulse
design algorithm, and is crucial for the development of that algorithm.
In our case, however, such splitting actually increases the number of
elementary computations:
when multiplying a vector, RF and transversal rotations require 9 and 4
multiplications, respectively, while direct multiplication by $R_t$ requires
only 9.
We therefore believe that the hard pulse approximation does not confer any
particular advantages on our approach.

Apart from the explicit Jacobians introduced here, additional steps may be taken
to reduce computation time.
Computation time is primarily determined by pulse length, and not on the grid
size (number of voxels) since computations are done voxel-wise
and can be easily parallelized to within GPU RAM limits.
Apart from increasing the simulation expense, longer pulses may also slow down
Algorithm~\ref{alg:AlterMin}, since we used L-BFGS for updating RF and
gradients.
In the future, to shorten the optimization time for online pulse design tasks,
it may be helpful to use coarser $\delta_t$ in the Bloch simulation
\cite{luo2021MultiScale} (here we used \SI{4}{\us} to match our scanner's
hardware dwell time), or parameterize the gradient waveforms to reduce the
optimization problem size (e.g., using B-splines as in \cite{Sun2016Joint}).

For the experiments presented, we used voxel resolution
\SI[product-units=power]{7.5x7.5x12}{\mm} and grid size \num{32 x 32 x 20} for
the pulse design.
For more complex target excitation patterns and/or a larger FOV (e.g., as in the
ISMRM parallel transmit pulse design challenge
\cite{Grissom2015PulseDesignChallenge}), it may be desirable to increase the
spatial resolution (maximum extent in excitation k-space) and/or grid size for
finer excitation accuracy control.
For instance, with a larger grid size, we would have space for in-plane ``don't
care'' region for the IV180M experiment, which may help reduce excitation error.
A larger grid size will increase the memory usage in simulation, for which the
use of multiple graphics cards may be needed to parallelize simulations across
voxels.

In the OV90 experiment (Figs. \ref{fig:OV90_pulse}--\ref{fig:OV90_result}), we
were able to apply VERSE \cite{Conolly1988Variable, Hargreaves2004Variable} to
the pulse designed with the small-tip approach \cite{Sun2016Joint} to avoid
violating the RF amplitude limit (after scaling to attain \ang{90} flip angle in
the OV).
However, this adjustment was possible only because the \SI{6.5}{\ms} pulse
happened to exceed peak RF only near the end of the pulse, allowing us to apply
the VERSE strategy in a relatively straightforward way.
In the more general case, where peak RF is exceeded during the middle of the
pulse, it is more difficult to apply the VERSE technique to 3D RF pulses such as
those designed here.
We found empirically that in the OV90 experiment, shorter pulses designed with
the small-tip approach tended to exceed peak RF during one or more intermediate
intervals (after scaling), and that we were therefore unable to carry out an
effective experimental evaluation for the purposes of the comparison presented
here (Figs.~\ref{fig:OV90_pulse}--\ref{fig:OV90_result}).
The proposed approach avoids this difficulty because peak RF is constrained as
described in \ref{sec:constraints};
our approach was in fact able to design a shorter (\SI{4}{\ms}) OV saturation
pulse with the same excitation error as in Fig.~\ref{fig:OV90_result} (not
shown).
In future work, a design approach that integrates VERSE into our method may be
useful to further shorten a pulse for a given excitation objective.

Pulse design problems are in general non-convex in terms of $b$ and $g$.
Due to a lack of theoretical tools for non-convex problem convergence analysis,
it is unclear how to best design an optimization algorithm for such problems
\textit{a priori}.
In Algorithm~\ref{alg:AlterMin}, instead of simultaneously updating both RF and
gradient waveforms, we chose to update them alternatingly as often done in
existing small-tip joint designs
\cite{Sun2016Joint,Cao2016Joint,Ma2011Joint,Yip2007Joint}.
In supplemental Figs. S6-S7, we compared the alternating scheme with the
simultaneous scheme, and found empirically that the alternating scheme optimizes
faster than the simultaneous scheme for the specific problem settings we have in
this work.
Unfortunately, the non-convexity prevents us from fully comprehending this
behavior, and we make no claims that the alternating approach used here is
optimal over the many possible alternatives.
Iteration stopping criteria for updating $b$ and $g$ are also commonly chosen
\textit{ad hoc}.
Besides limiting the maximum number of iterations as we have done in
Algorithm~\ref{alg:AlterMin}, another option can be setting a threshold to
assert large-enough loss decreases and/or updates of the variables at each
iteration.
However, due to the $\tan^{-1}$ change of variables, a minuscule update of
$b$ and $s$ near their limits will be mapped to a vast difference in the
optimization variables $\tilde{\rho}$ and $\tilde{s}$, respectively.
As an alternative to the updates of variables, one can threshold the norms of
variable derivatives, or the change of $b$ and $s$ as the iteration stopping
criteria.

Our approach may remind readers of optimal control (OC) based pulse design
methods \cite{Conolly1986Optimal, Grissom2009Fast}.
Comparing the OC formulation with our approach, we can make the following
observations (ignoring the penalization and relaxation terms):
Eq. \eqref{eq:iterforw} is the forward propagation of spin states in OC (state
equation);
the first identity in Eq. \eqref{eq:iterback} is the backward propagation of OC
Lagrange multiplier (costate equation);
the second identity in Eq. \eqref{eq:iterback} is the derivative for iteratively
optimizing B-effective as the control.
Being one step in the computation of excitation losses, our auto-differentiable
Bloch simulator enables reusing the forward and backward iterations regardless
of the actual design loss function chosen, and propagating the derivatives to
the actual controls, i.e., the RF and gradient waveforms.
In future works, it may be beneficial to employ tools from control research.
For instance, optimization algorithms from OC may accelerate or replace
algorithm~\ref{alg:AlterMin}.

Like many other pulse design works \cite{Sun2016Joint, Ma2011Joint,
Cao2016Joint, Yip2007Joint}, the experiments we presented in this
work assumes that the off-resonance map is known.
We have not attempted to enforce robustness to unknown off-resonance patterns,
however the auto-differentiation nature of this work allows incorporating such
robustness into the loss as an error metric or regularization for the joint
design.
In particular, noting the relation of our tool to the control framework, we
anticipate that incorporation of robust control methods can improve robustness
to off-resonance errors.

A major advantage of our approach is that it enables designs involving arbitrary
loss functions, enabling novel design formulations that have so far not been
tractable.
For example, we demonstrated in \eqref{eq:loss_IV180} a loss involving only
longitudinal magnetization.
Other possibilities may include the addition of constraints or regularization
terms involving specific absorption rate (SAR) or peripheral nerve stimulation
(PNS).
Another important feature is that the method back-propagates derivatives
throughout the Bloch simulator,
which may facilitate development of neural network based pulse design
approaches.

A limitation of our method is that it only works for fixed pulse length, as
determined by the initial waveforms.
As shown in the pulse plots in Figs. \ref{fig:OV90_pulse},
\ref{fig:IV180_result}, and \ref{fig:IV180M_result}, there are temporal
intervals where neither the RF, gradients, nor slew rates are hitting their
constraints.
This may suggest that the pulses can be shortened without sacrificing excitation
accuracy.
Pulse shortening can be formed as a minimum-time pulse design problem
\cite{Conolly1986Optimal} in the OC context.
Noting the relation of our method to the OC approach, for future work, we expect
employing existing OC tools to be helpful in overcoming the fixed length
limitation.

\section{Conclusion}\label{sec:conc}
In this work, we have proposed a novel approach based on auto-differentiation
tools for the joint design of RF and gradient waveforms, and validated it with
multi-dimensional spatially tailored excitation tasks in MRI.
Using short (\SI{<5}{\ms}) excitation pulses and single (body) coil RF
transmission, we demonstrated experimentally that even a fairly complex 3D
spatial pattern (block-M) can be selectively inverted.
Our method is not limited to specific design objectives.
To reduce computation time and memory requirements, we derived explicit
Jacobians for the Bloch simulator, as the simulation steps are typically the
most computationally demanding.
We used a change of variables to enforce hardware limits, enabling use of
simpler unconstrained optimization.
We anticipate that the proposed method will be useful for a broad range of
excitation pulse design problems in MRI.

\appendices

% \section*{Appendix}

% \section*{Acknowledgment}

\bibliographystyle{IEEEtran}
\bibliography{AutoDiffPulse_bib}

\appendices

% \section*{Appendix}

% \section*{Acknowledgment}
\onecolumn
\setcounter{figure}{0}
\renewcommand{\thefigure}{S\arabic{figure}}
\section{Additional simulation results}

For completeness, we present here both the initial pulses designed as described
in III-C in the main text, and the corresponding optimized pulses obtained with
the proposed approach.
In the case of OV90, we also include the small-tip pulse.
In each case, we show the simulated excitation pattern for each pulse.

In addition to IV90, IV180, IV180M, we include here a cuboid IV inversion
pulse based on the B0 field map acquired in the brain of a healthy volunteer.
This is done to demonstrate the feasibility of designing an IV inversion pulse
with the proposed approach using a more realistic B0 map than that shown in
Fig.~3.
In addition, for that simulation experiment we compare the optimized pulse with
a pulse obtained by only optimizing the RF waveform, i.e., keeping the gradient
waveforms fixed at their initial shapes.
This is done to assess the relative importance of also optimizing the gradient
waveforms.

The key takeaways from these figures are:
(1) The excitation patterns produced by the initial pulses are substantially
inferior to the optimized patterns.
(2) The initial and optimized excitation k-space trajectories tend to be similar,
suggesting that a local minimum is obtained.
(3) The initial and optimized RF waveforms (amplitude and phase), on the other
hand, differ markedly from each other, suggesting relatively weak dependence on
initial RF waveform.
This may be due to the fact that each RF sample can be optimized independently
of the other samples, unlike gradient waveforms that are subject to slew rate
constraints.
(4) Despite the similarity between the initial and optimized gradient waveforms,
the optimized waveforms produce a more accurate excitation than the pulse
obtained by optimizing only the RF waveform (Fig. \ref{fig_supp:IV180_human}).
\vspace{-\baselineskip}

\subsection{OV90}

% \clearpage
\begin{figure*}[!h]
  \centering
  \includegraphics[width=0.82\textwidth]{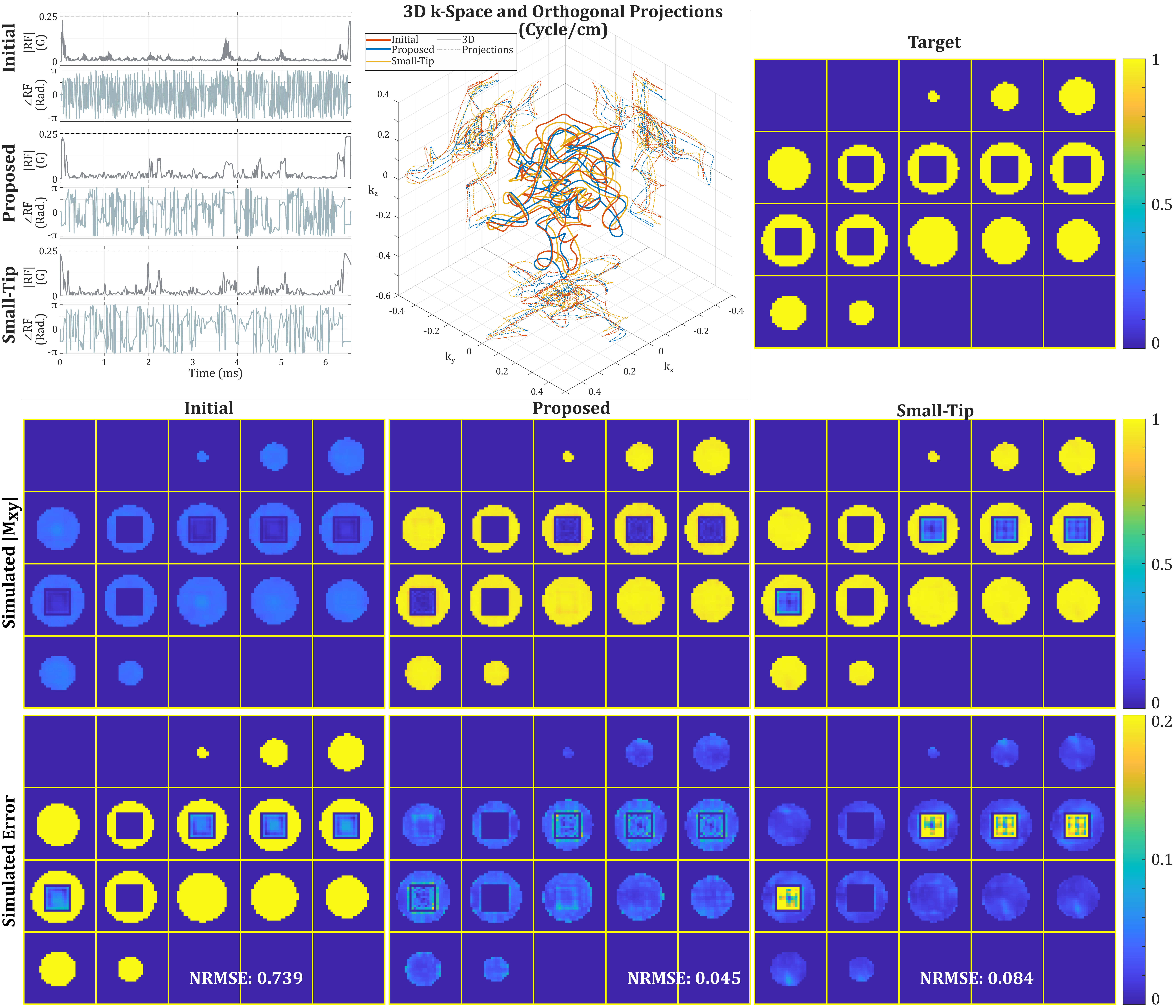}
  % \vspace{-\baselineskip}
  \caption{
    OV saturation pulses (and results) for the cuboid IV,
    designed with our approach (Proposed) and Sun's small-tip method
    (experiment OV90 in the main manuscript).
    On the 3D k-space plot, the three trajectories explore largely overlapping
    regions in excitation k-space.
    Our approach has much smaller (-46\%) NRMSE in simulation compared to
    Small-Tip.
    Small-Tip approach has larger error inside the IV:
    This is expected, as the method produces only small-tip pulses, that we then
    scaled to meet the large-tip objective.
    The scaling increases excitation error inside IV while reducing error in the
    OV.
    Our approach directly designs large-tip pulses without this type of
    `scaling' error.
  }
  \vspace{-\baselineskip}
  \label{fig_supp:OV90}
\end{figure*}

\clearpage
\subsection{IV180}

\vspace{\baselineskip}

\begin{figure*}[!h]
  \centering
  \includegraphics[width=\textwidth]{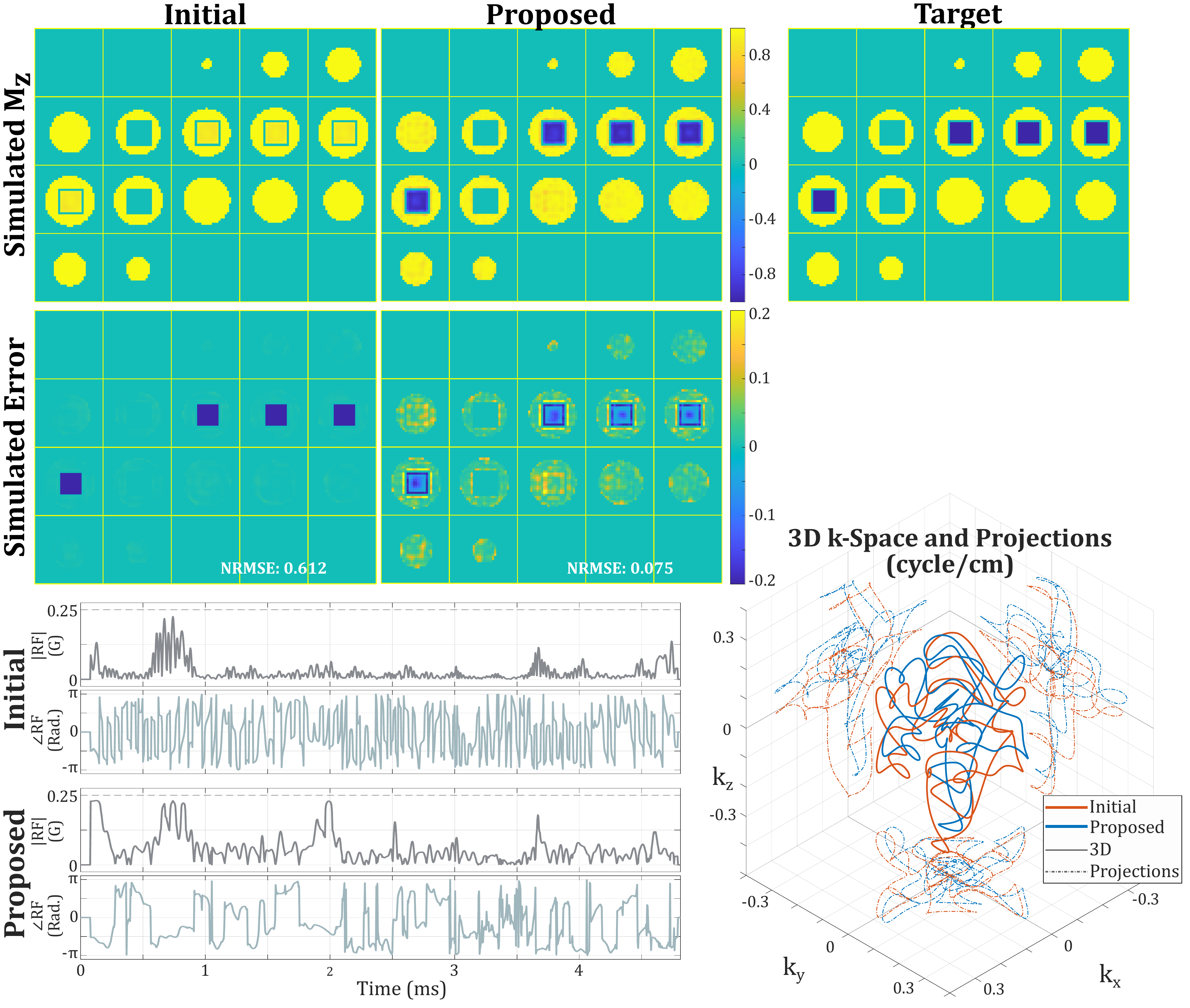}
  \vspace{-\baselineskip}
  \caption{
    Cuboid IV inversion (experiment IV180 in the main manuscript).
    The lower left panel compares the RF waveforms of the initial and optimized
    (Proposed) pulses.
    The lower right panel compares the k-space trajectories of the two pulses.
  }
  \vspace{-\baselineskip}
  \label{fig_supp:IV180}
\end{figure*}

\clearpage

\subsection{IV180M}

\vspace{\baselineskip}

\begin{figure*}[!h]
  \centering
  \includegraphics[width=\textwidth]{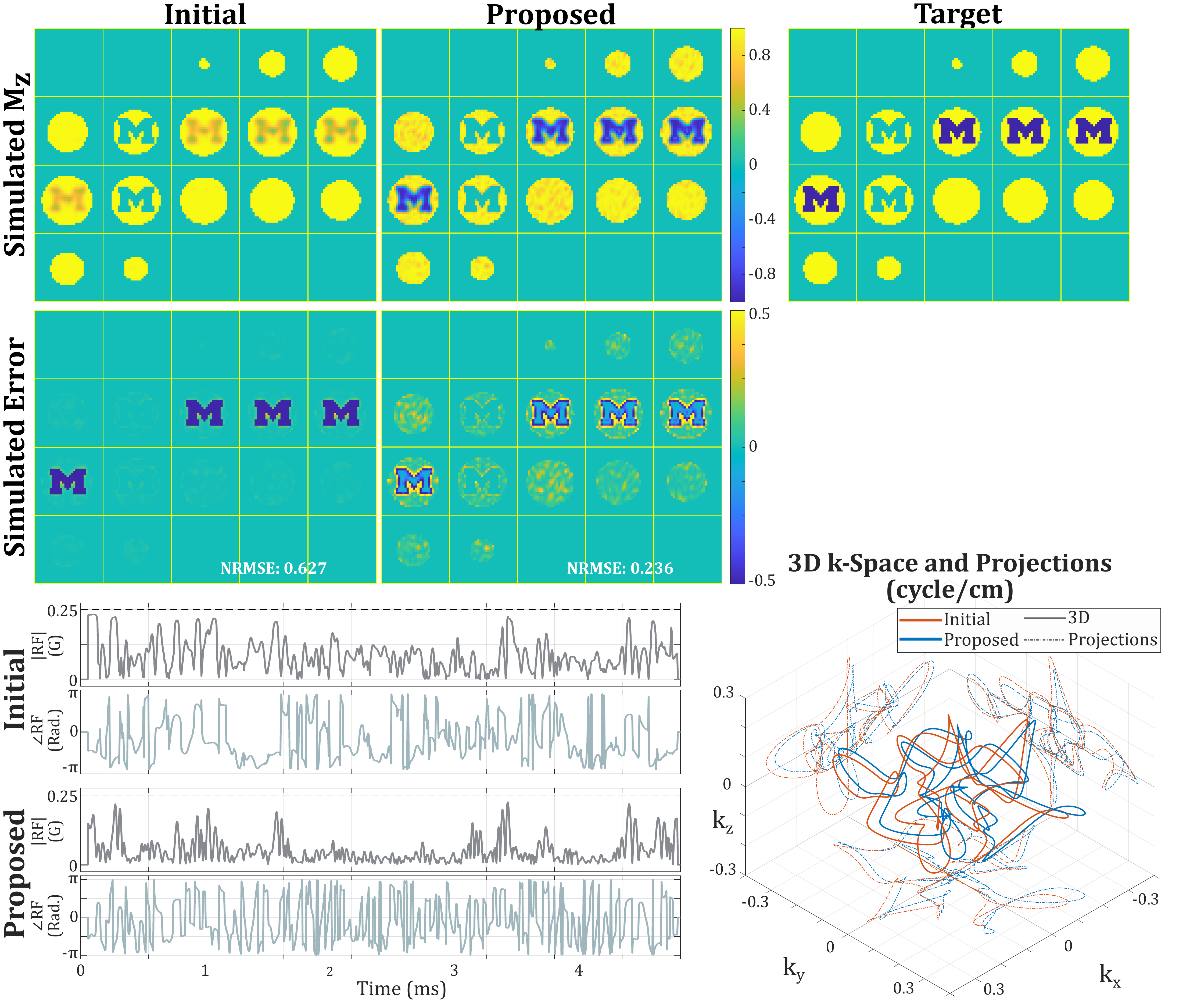}
  \vspace{-\baselineskip}
  \caption{
    Block-M IV inversion pulse (experiment IV180M in the main manuscript).
    The lower left panel compares the RF waveforms of the initial and optimized
    (Proposed) pulses.
    The lower right panel compares the k-space trajectories of the two pulses.
  }
  \vspace{-\baselineskip}
  \label{fig_supp:IV180M}
\end{figure*}

\clearpage

\subsection{Brain Simulation}
% \vspace{-\baselineskip}
\begin{figure*}[!h]
  \centering
  \includegraphics[width=0.9\textwidth]{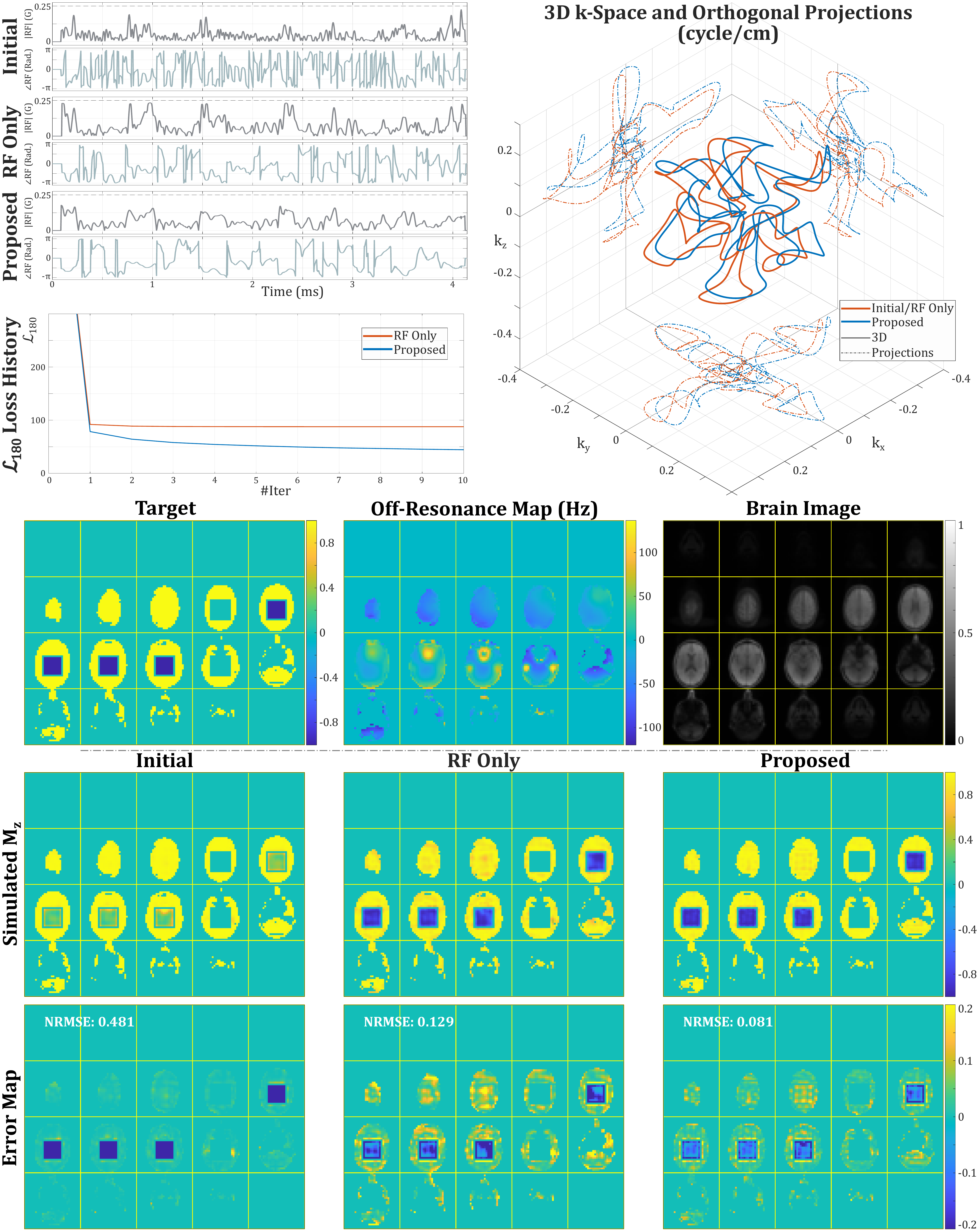}
  \vspace{-0.5\baselineskip}
  \caption{
    Cuboid IV inversion based on a B0 (off-resonance) map obtained in a
    volunteer.
    Three pulses are compared:
    (1) Initial,
    (2) 'RF Only', obtained by keeping the gradients fixed at their initial
    shapes and optimizing only the RF waveform with the proposed
    auto-differentiation approach,
    and (3) the proposed jointly optimized pulse.
    While our optimized k-space trajectory is similar to the initial k-space
    trajectory, the jointly optimized pulse (Proposed) attains an excitation
    accuracy (NRMSE: 8.1\%) that is 37\% better than the 'RF Only' pulse (NRMSE:
    12.9\%).
    This improvement is also reflected in the convergence ($\mathcal{L}_{180}$
    loss history) plot.
  }
  \label{fig_supp:IV180_human}
\end{figure*}
\clearpage

\section{Unnormalized Inversion Images}
\begin{figure*}[!h]
  \centering
  \includegraphics[width=0.9\textwidth]{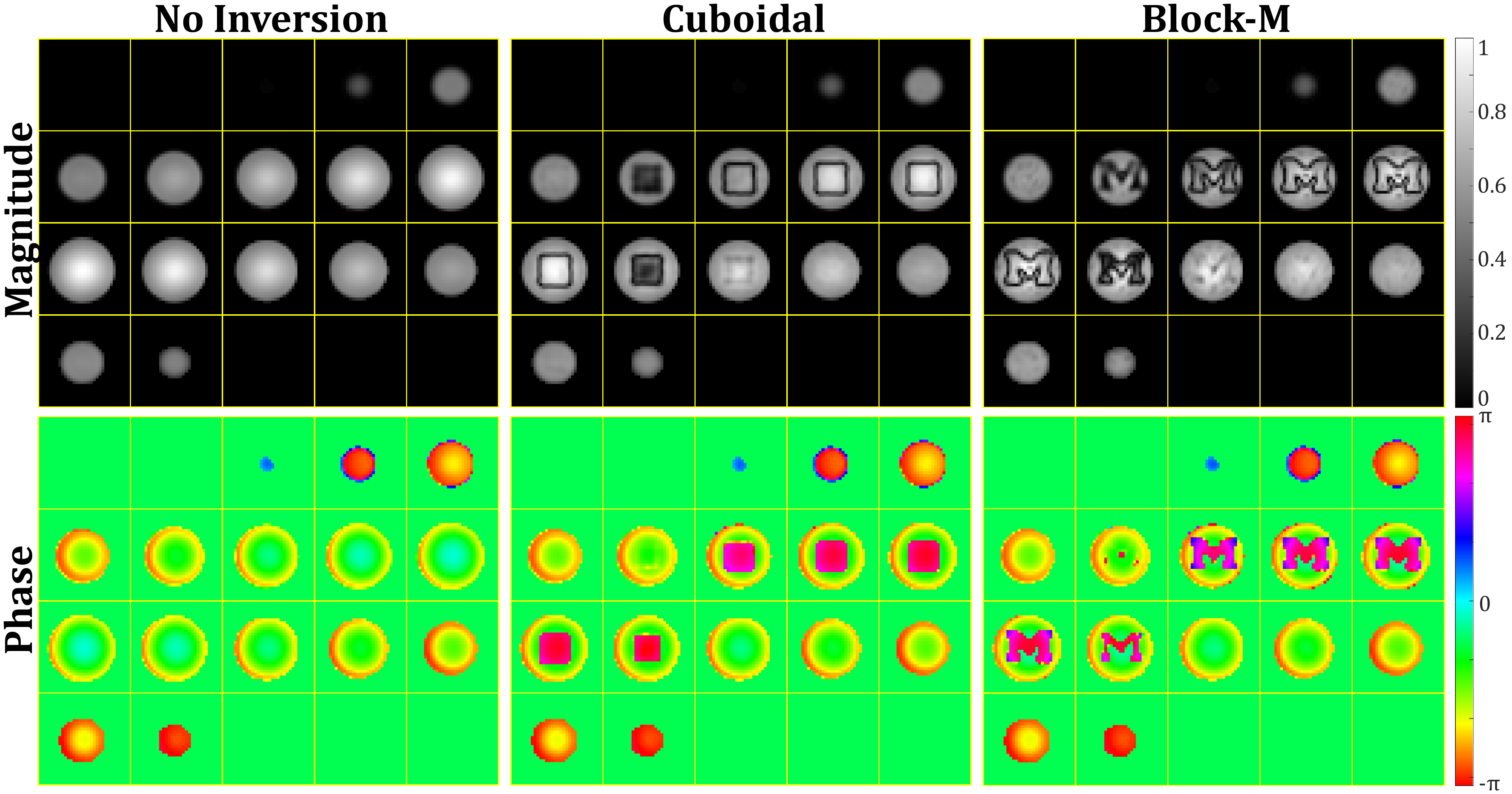}
  \vspace{-0.5\baselineskip}
  \caption{
    Images from the IV180 and IV180M experiments.
    ``Cuboidal'' and ``Block-M'' are the raw images of the IV180 and IV180M
    experiments, respectively.
    ``No Inversion'' was acquired using the same sequence (TR/TE, flip-angle,
    readout trajectory), except the inversion pulse in the sequence has RF
    amplitude set to 0.
    The 3 sets of magnitude and phase images shown here share the same image
    intensity and phase variations due to receiver coil sensitivity.
  }
  \label{fig_supp:normal}
\end{figure*}

In Fig. \ref{fig_supp:normal}, we show the unnormalized images for the 3D
spatially tailored inversion experiments (IV180, IV180M).
In the main text, we normalized the ``Cuboidal'' and ``Block-M'' images by
element-wise division by the ``No Inversion'' image, which eliminates the
image intensity and phase variations due to receiver coil sensitivity.

\clearpage

\section{Alternating vs Simultaneous Minimization}
Here we compare the alternating optimization used in the main text with a
simultaneous update scheme that optimizes $b$ (RF waveform) and $g$ (the three
gradient waveforms) together at each iteration rather than fixing one and
optimizing the other.
We observe empirically that for the $\mathcal{L}_{90}$ and $\mathcal{L}_{180}$
losses defined in the main text, with extended kt-points initializations, the
alternating update decreases the design losses faster than the simultaneous
update.
However, the two objectives are both non-convex in terms of $b$ and $g$, which
makes this behavior difficult to analyze.
We therefore cannot claim that this alternating scheme will outperform
simultaneous updates in the general case (i.e., for all other design problems).

\begin{figure*}[!h]
  \centering
  \includegraphics[width=0.75\textwidth]{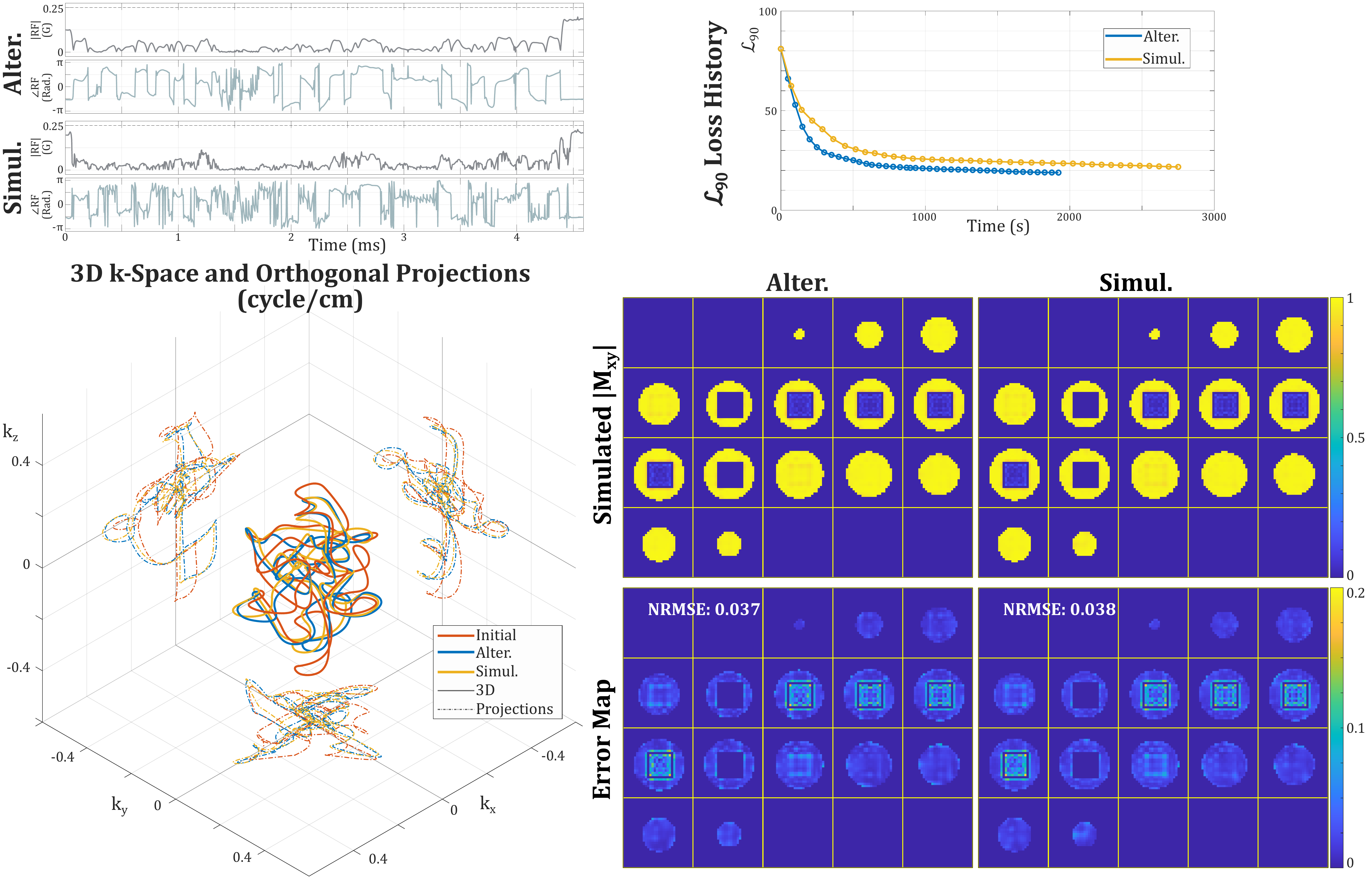}
  \vspace{-0.5\baselineskip}
  \caption{
    Comparison of alternating (Alter.) and Simultaneous (Simul.) minimization.
    The two approaches find similar but different local minima according to the
    RF and k-space plots (Left Panel).
    On the Top Right, the loss for the first 40 iterations is plotted.
    The computation time for each iteration is slightly longer for the
    simultaneous L-BFGS updates.
    The simultaneous updating scheme converges slightly slower, while the
    eventual excitation performance of the two schemes is comparable (Lower
    Right).
  }
  \label{fig_supp:altersimulOV90}
\end{figure*}

\begin{figure*}[!h]
  \centering
  \includegraphics[width=0.75\textwidth]{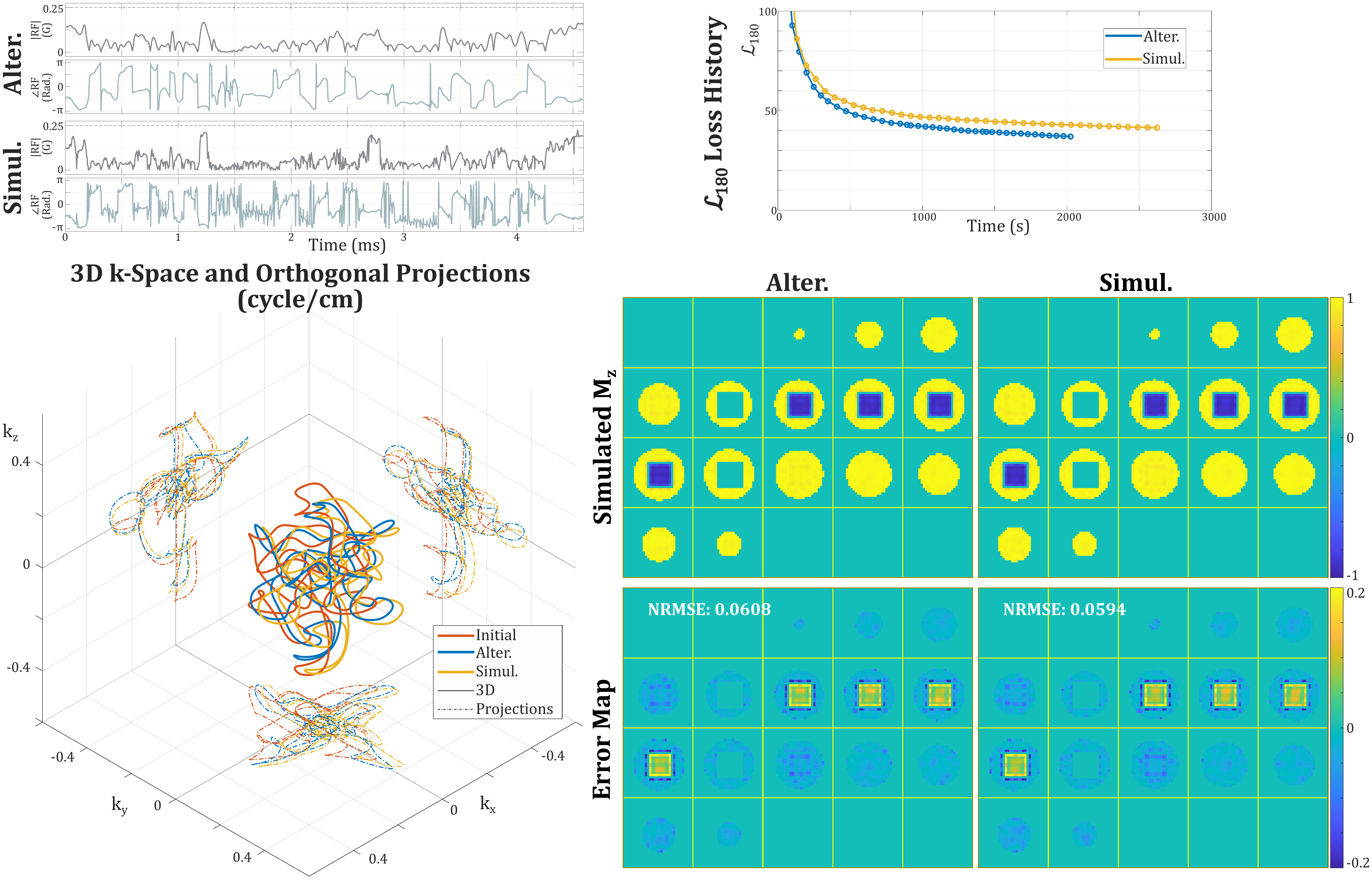}
  \vspace{-0.5\baselineskip}
  \caption{
    Same comparison as in Fig. \ref{fig_supp:altersimulOV90}, for the IV180
    design.
    The simultaneous approach attains a slightly better inversion, but ends with
    a higher loss value (likely due to the RF power penalization term).
  }
  \label{fig_supp:altersimulIV180}
\end{figure*}
\clearpage

\section{Impact of a Small Gradient Delay}
\begin{figure*}[!h]
  \centering
  \includegraphics[width=\textwidth]{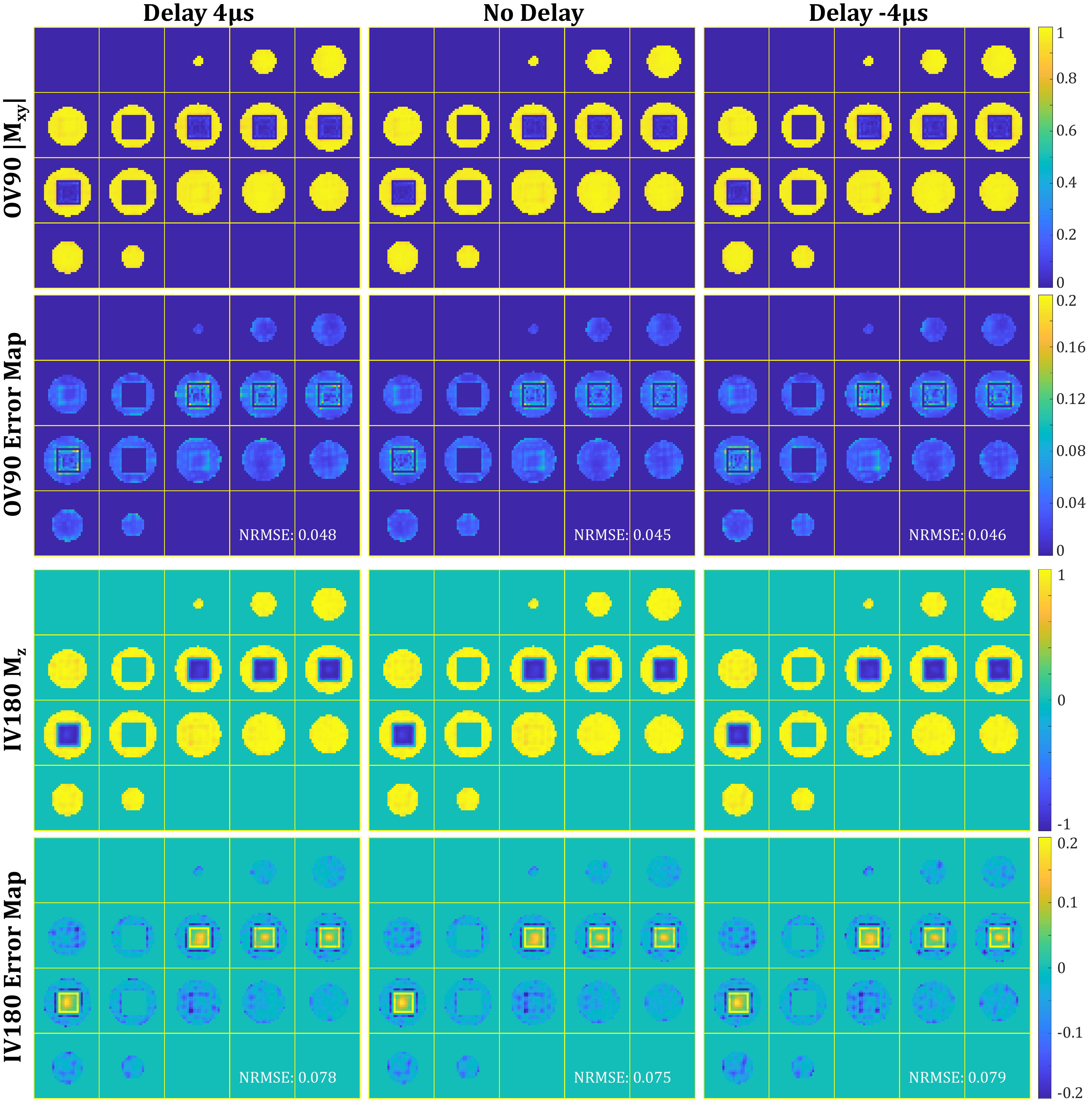}
  \vspace{-0.5\baselineskip}
  \caption{
    Simulated excitation performances of our designed OV90 and IV180 pulses
    under different delays of the applied gradient fields.
  }
  \label{fig_supp:delay}
\end{figure*}

On modern MRI scanners, the physically realized gradient waveforms are typically
slightly misaligned in time relative to the RF waveform, even after the vendor's
built-in gradient delay correction is applied.
This delay is on the order of the gradient sampling (dwell, or raster) time,
which on our scanner is \SI{4}{\us}.
To assess robustness against such delays, we simulated the excitation produced
by the OV90 and IV180 pulses for delays of \SI{4}{\us} and \SI{-4}{\us}.
As shown in Fig. \ref{fig_supp:delay}, such delays led to excitation patterns
that are visually nearly indistinguishable from the original patterns with no
delay, and degraded the performance of our designed pulses by less than \num{1}
percentage point in NRMSE.

\clearpage

\section{Impact of Incorrect Off-Resonance}
\begin{figure*}[!h]
  \centering
  \includegraphics[width=\textwidth]{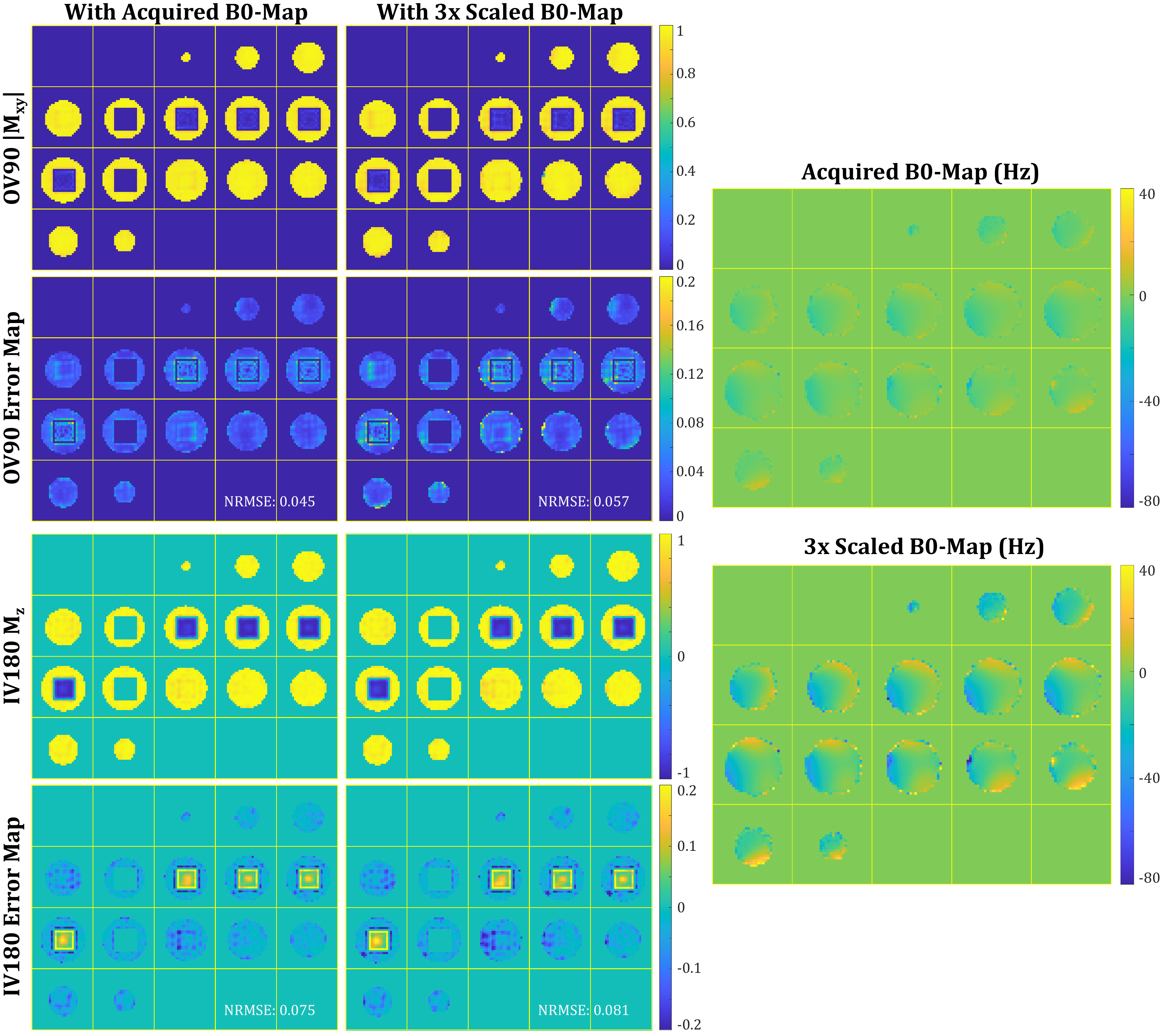}
  \vspace{-0.5\baselineskip}
  \caption{
    Simulated excitation performances of our designed OV90 and IV180 pulses
    under different off-resonance maps.
  }
  \label{fig_supp:offres}
\end{figure*}

Motion or respiratory effects can cause mismatch between the acquired and actual
off-resonance patterns.
To assess robustness against such mismatch, we simulated the excitation produced
by the same OV90 and IV180 pulses under different off-resonance maps: (i) the
acquired off-resonance map, and (ii) the acquired off-resonance map scaled by a
multiplicative factor of 3.
As shown in Fig. \ref{fig_supp:offres}, such mismatch led to a very similar
excited patterns with about \num{1} percentage point increases in NRMSE.

\clearpage

\end{document}